\documentclass[10pt,twocolumn,letterpaper]{article}
\PassOptionsToPackage{table}{xcolor}
\usepackage{iccv}
\usepackage{times}
\usepackage{epsfig}
\usepackage{graphicx}
\usepackage{amsmath}
\usepackage{amssymb}
\usepackage{multirow}
\definecolor{iccvblue}{rgb}{0.21,0.49,0.74}
\usepackage[pagebackref,breaklinks,colorlinks,allcolors=iccvblue]{hyperref}

\usepackage{xcolor}
\usepackage{wrapfig}
\usepackage{soul}

\definecolor{lightgray}{rgb}{0.4, 0.4, 0.4}
\definecolor{yellow}{rgb}{1,1, 0.6}
\definecolor{lightyellow}{rgb}{1,1, 0.8}
\definecolor{orange}{rgb}{1, 0.8, 0.6}
\definecolor{coral}{RGB}{246,131,65}
\definecolor{pinkred}{rgb}{1, 0.6, 0.6}
\definecolor{hotpink}{RGB}{238,64,195}
\definecolor{lavender}{RGB}{207,226,243}
\definecolor{gainsboro}{RGB}{208,224,227}
\definecolor{gainsboro2}{RGB}{217,234,211}
\definecolor{blanchedalmond}{RGB}{252,229,205}

\definecolor{lightgreen}{rgb}{0.88, 1.0, 0.88}
\definecolor{lightpink}{rgb}{1.0, 0.88, 0.88}
\definecolor{lightyellow}{rgb}{1.0, 1.0, 0.88}

\usepackage{datenumber}
\usepackage{calc}
\usepackage[mmddyyyy]{datetime}
\newcounter{datetoday}
\newcounter{diffyears}
\newcounter{diffmonths}
\newcounter{diffdays}
\newcommand{\difftoday}[3]{%
      \setmydatenumber{datetoday}{\the\year}{\the\month}{\the\day}%
      \setmydatenumber{diffdays}{#1}{#2}{#3}%
      \addtocounter{diffdays}{-\thedatetoday}%
      \ifnum\value{diffdays}>0
        \def\diffbefore{}%
        \def\diffafter{left}%
      \else
        \def\diffbefore{}%
        \def\diffafter{ago}%
        \setcounter{diffdays}{-\value{diffdays}}%
      \fi
      \setcounter{diffyears}{\value{diffdays}/365}%
      \setcounter{diffdays}{\value{diffdays}-365*\value{diffyears}}%
      \setcounter{diffmonths}{\value{diffdays}/30}%
      \setcounter{diffdays}{\value{diffdays}-30*\value{diffmonths}}%
      \diffbefore
      \ifnum\value{diffyears}=0
      \else
        \ifnum\value{diffyears}>1
            \thediffyears\space years,
        \else
            \thediffyears\space year,
        \fi
      \fi
      \ifnum\value{diffmonths}=0
      \else
        \ifnum\value{diffmonths}>1
            \thediffmonths\space months
        \else
            \thediffmonths\space month
        \fi
      \fi
      \ifnum\value{diffdays}=0
      \else
        \ifnum\value{diffdays}>1
            \thediffdays\space days
        \else
            \thediffdays\space day
        \fi
      \fi
      \diffafter
}

\newcommand{\IGNORE}[1]{}

\definecolor{darkred}{rgb}{0.6,0,0}

\usepackage[utf8]{inputenc}   
\usepackage[T1]{fontenc}      
\usepackage{kotex}            
\usepackage{booktabs} 
\usepackage{amsbsy}
\usepackage[mathscr]{euscript}
\usepackage{amsmath}
\usepackage{stackrel}
\usepackage{nicefrac}
\usepackage{kotex}
\usepackage{mathbbol}

\usepackage{dblfloatfix}
\usepackage{float}

\usepackage{subcaption}
\usepackage{pifont}  
\usepackage{colortbl}
\usepackage{array}
\usepackage[accsupp]{axessibility}
\definecolor{lightgreen}{rgb}{0.88, 1.0, 0.88}
\definecolor{lightpink}{rgb}{1.0, 0.88, 0.88}
\definecolor{lightyellow}{rgb}{1.0, 1.0, 0.88}

\newcommand{\cmark}{\textcolor{green}{\ding{51}}}  
\newcommand{\xmark}{\textcolor{red}{\ding{55}}}    

\begin{document}
\title{A Real-world Display Inverse Rendering Dataset}

\author{Seokjun Choi\footnotemark[1] ~ ~ ~
Hoon-Gyu Chung\footnotemark[1] ~ ~ ~
Yujin Jeon\footnotemark[1] ~ ~ ~
Giljoo Nam\footnotemark[2] ~ ~ ~
Seung-Hwan Baek\footnotemark[1] \\
\footnotemark[1]~ POSTECH   ~ ~ ~   ~ ~ ~  \footnotemark[2]~ Meta\\
}
\maketitle

\begin{abstract}
\vspace{-4mm}

{
Inverse rendering aims to reconstruct geometry and reflectance from captured images. Display-camera imaging systems offer unique advantages for this task: each pixel can easily function as a programmable point light source, and the polarized light emitted by LCD displays facilitates diffuse-specular separation. Despite these benefits, there is currently no public real-world dataset captured using display-camera systems, unlike other setups such as light stages. 
This absence hinders the development and evaluation of display-based inverse rendering methods.
In this paper, we introduce the first real-world dataset for display-based inverse rendering. To achieve this, we construct and calibrate an imaging system comprising an LCD display and stereo polarization cameras. We then capture a diverse set of objects with diverse geometry and reflectance under one-light-at-a-time (OLAT) display patterns. We also provide high-quality ground-truth geometry. Our dataset enables the synthesis of captured images under arbitrary display patterns and different noise levels. Using this dataset, we evaluate the performance of existing photometric stereo and inverse rendering methods, and provide a simple, yet effective baseline for display inverse rendering, outperforming state-of-the-art inverse rendering methods.
Code and dataset are available on our project page at \href{https://michaelcsj.github.io/DIR/}{https://michaelcsj.github.io/DIR/}.
}
\end{abstract}

\section{Introduction}
\label{sec:intro}
Inverse rendering is a long-standing problem in computer vision and graphics, aiming to recover scene properties such as geometry and reflectance from captured images~\cite{lensch2003image,ramamoorthi2001signal}. Recent progress in inverse rendering methods heavily rely on datasets that provide images of objects under well-characterized multiple lighting conditions~\cite{bi2024gs3,chung2024differentiable,saito2024relightable}, allowing for evaluation and training of models that infer geometry and reflectance from images.

Among various inverse rendering setups, display-camera imaging systems offer unique advantages. Unlike conventional light stages~\cite{ghosh2009estimating, kampouris2018diffuse,ma2007rapid,sato2003appearance}, displays can serve as high-resolution, programmable light sources, allowing convenient control over illumination~\cite{aittala2013practical, zhang2023deep}. Moreover, LCD displays emit polarized light, which facilitates the separation of diffuse and specular reflections~\cite{choi2024differentiable,lattas2022practical}. These characteristics make display-camera systems a compelling choice for inverse rendering research. However, despite their potential, the lack of publicly available datasets captured using such systems has hindered progress in this direction. Unlike other setups, such as light stages, which have been extensively used for photometric stereo and reflectance capture, display-camera inverse rendering lacks a standardized benchmark for method development and comparison.

In this paper, we introduce the first real-world dataset for display-based inverse rendering. We construct a display-camera imaging system consisting of a LCD monitor and a stereo polarization camera setup, enabling controlled illumination capture at two views with diffuse and specular separation. Using this system, we capture a diverse set of objects with varying geometries and reflectance properties under one-light-at-a-time (OLAT) display patterns. Each object is accompanied by ground-truth geometry obtained via structured-light scanning, enabling precise evaluation of inverse rendering methods. Our dataset also supports synthetic relighting and noise simulation, allowing researchers to generate novel lighting conditions using linear combinations of captured images.
We also introduce a simple baseline method for display inverse rendering that effectively addresses associated challenges, outperforming previous methods. Our specific contributions are as follows:
\begin{itemize}
    \item We build and calibrate a display-camera imaging system incorporating display backlight, which enables display-based illumination and stereo polarization imaging.
    \item We acquire the first high-quality real-world dataset for display-camera inverse rendering, featuring objects with diverse reflectance and ground-truth geometry.
    \item We evaluate existing photometric stereo and inverse rendering methods on our dataset, highlighting the challenges of display inverse rendering.
    \item {We propose a simple yet effective baseline for display inverse rendering, outperforming previous methods.}
\end{itemize}

\section{Related Work}
\label{sec:related}

\paragraph{Imaging Systems for Inverse Rendering}
Inverse rendering typically requires observations of a target object under various lighting conditions. 
In the literature, different hardware configurations to modulate lighting conditions have been proposed.
Light stages, a dome structure equipped with numerous high brightness LEDs, offer dense light-view angular samples for high-quality inverse rendering at the cost of large form factors and high instrumentation costs~\cite{sato2003appearance, ma2007rapid, ghosh2009estimating, kampouris2018diffuse}.
Flash photography with mobile cameras provides a practical multi-view, multi-light setup, capturing many images from different views~\cite{nam2018practical, azinovic2023high, deschaintre2021deep, hui2017reflectance, riviere2014mobile}. 
However, this requires moving the cameras and capturing objects multiple times.
Using displays as controllable light sources provides a cost-effective and compact alternative, enabling convenient multi-light capture, having a potential for practical and high-quality inverse rendering~\cite{aittala2013practical, lattas2022practical, zhang2023deep, choi2024differentiable}. 
Display-camera systems present unique challenges and opportunities due to near-field lighting effects, limited light power, polarization properties of LCDs, and constrained light-view angular sampling. Addressing these challenges is an open problem.

\paragraph{Inverse Rendering Dataset}
Table~\ref{tab:dataset} summarizes representative publicly available datasets for inverse rendering. While synthetic datasets provide ground truth under ideal scenarios~\cite{chen2018ps,ikehata2018cnn,ikehata2022universal}, real-world datasets offer environments for realistic evaluation. Existing real-world datasets are captured with various imaging systems such as commodity cameras~\cite{grosse2009ground,kuang2022neroic,rudnev2022nerf}, light probes~\cite{kuang2023stanfordorb}, gantries~\cite{ren2022diligent102,wang2023diligent,guo2024diligenrt}, robots~\cite{jensen2014large,toschi2023relight}, and light stages~\cite{chabert2006relighting,yang2024acquisition,liu2024openillumination}. 
Despite the increasing availability of real-world datasets, existing datasets fail to comprehensively evaluate inverse rendering in display-camera settings due to the use of other imaging systems for data acquisition. 
{Recently, Choi et al.~\cite{choi2024differentiable} employs 3D-printed objects for display photometric stereo. However, the 3D-printed dataset has limited material diversity, unsuitable as an inverse rendering dataset for real-world diverse objects.}

\begin{table}[t]

        \centering
        \setlength{\tabcolsep}{2pt}
        \caption{ \label{tab:dataset} \textbf{Real-world inverse rendering datasets.} We present the first dataset for display inverse rendering with calibrated display and stereo polarization cameras. We also provide high-quality ground-truth geometry.}
        \scalebox{0.8}{%
        \begin{tabular}{c|c|c|c|c}
            \hline
             \multirow{2}{*}{Dataset} & Illumination  & Illumination &  Ground-truth& \multirow{2}{*}{Polarization}  \\
              & system & type & geometry & \\ 
        \toprule[1pt]
            Alldrin et al.~\cite{alldrin2008photometric} & Light rig  & Far-field & \xmark & \xmark                               \\
            Grosse et al.~\cite{grosse2009ground} & Light rig & Far-field & \xmark  & \xmark                          \\
            Xiong et al.~\cite{xiong2014shading} & Light rig & Far-field & \xmark  & \xmark                             \\
            Jensen et al.~\cite{jensen2014large} & Light rig & Far-field & \cmark  & \xmark                                        \\
            Shi et al.~\cite{shi2016benchmark} & Light rig& Far-field & \cmark  & \xmark                             \\
            Li et al.~\cite{Li2020diligentmv} & Light rig& Far-field  & \cmark & \xmark                \\
            Mecca et al.~\cite{mecca2021luces} & Light rig& Near-field & \cmark  & \xmark                              \\ 
            Chabert et al.~\cite{chabert2006relighting}& Light stage & Far-field & \xmark  & \xmark                          \\
            Liu et al.~\cite{liu2024openillumination} & Light stage & Far-field & \xmark  & \xmark                            \\
            Yang et al.~\cite{yang2024acquisition} & Light stage & Far-field & Pseudo  & \cmark                        \\
            Toschi et al.~\cite{toschi2023relight} & Gantry & Far-field & \xmark & \xmark                                \\
            Kuang et al.~\cite{kuang2022neroic} & In-the-wild & Env. map & \xmark  & \xmark                                                \\
            Kuang et al.~\cite{kuang2023stanfordorb} & In-the-wild & Env. map & \cmark & \xmark                                                  \\ 
        \bottomrule[1pt]
            {Ours} & LCD display & Near-field & \cmark  & \cmark         \\ \hline
            
        \end{tabular}
        }
    \label{tab:pattern_photometric_stereo}
\end{table}

\paragraph{Inverse Rendering Methods}
Learning-based inverse-rendering methods utilize CNN~\cite{li2020inverse, li2018learning, boss2020two, sang2020single, wang2021learning, sengupta2019neural,wei2020object, yu2019inverserendernet}, RNN~\cite{lichy2021shape}, transformers~\cite{ikehata2023scalable, zhu2022irisformer}, and diffusion models~\cite{chen2025intrinsicanything, sartor2023matfusion, lyu2023diffusion,litman2024materialfusion,enyo2024diffusion,liang2025diffusionrenderer,he2024neural} to infer geometry and reflectance in a data-driven manner. 
In contrast, analysis-by-synthesis methods take a physics-based approach, iteratively optimizing geometry and reflectance, ensuring that rendered images match the input images via differentiable forward rendering. Various differentiable rendering techniques have been explored, including volumetric rendering~\cite{zhang2021nerfactor, zhang2021physg, zhang2022modeling, zhang2022iron, liu2023nero, wang2024nep, yang2022ps, zeng2023relighting}, spherical Gaussians~\cite{zhang2021physg, zhang2022modeling}, tensor-based formulations~\cite{jin2023tensoir}, point-based rendering~\cite{chung2024differentiable}, and Gaussian-based representations~\cite{chung2024interpretable, liang2024gs, jiang2024gaussianshader, gao2023relightable, bi2024gs3}, and image-based neural representations~\cite{li2022neural}.
Inverse rendering for display-camera systems introduces unique challenges and benefits for reconstruction methods due to near-field lighting conditions, display backlight, low signal-to-noise ratios, LCD polarization effects, and non-uniform angular sampling~\cite{aittala2013practical, lattas2022practical, zhang2023deep}. Developing reconstruction methods for display inverse rendering remains for future research.

\section{Display-camera Imaging System}

\begin{figure}[htbp]
    \centering
    \includegraphics[width=\linewidth]{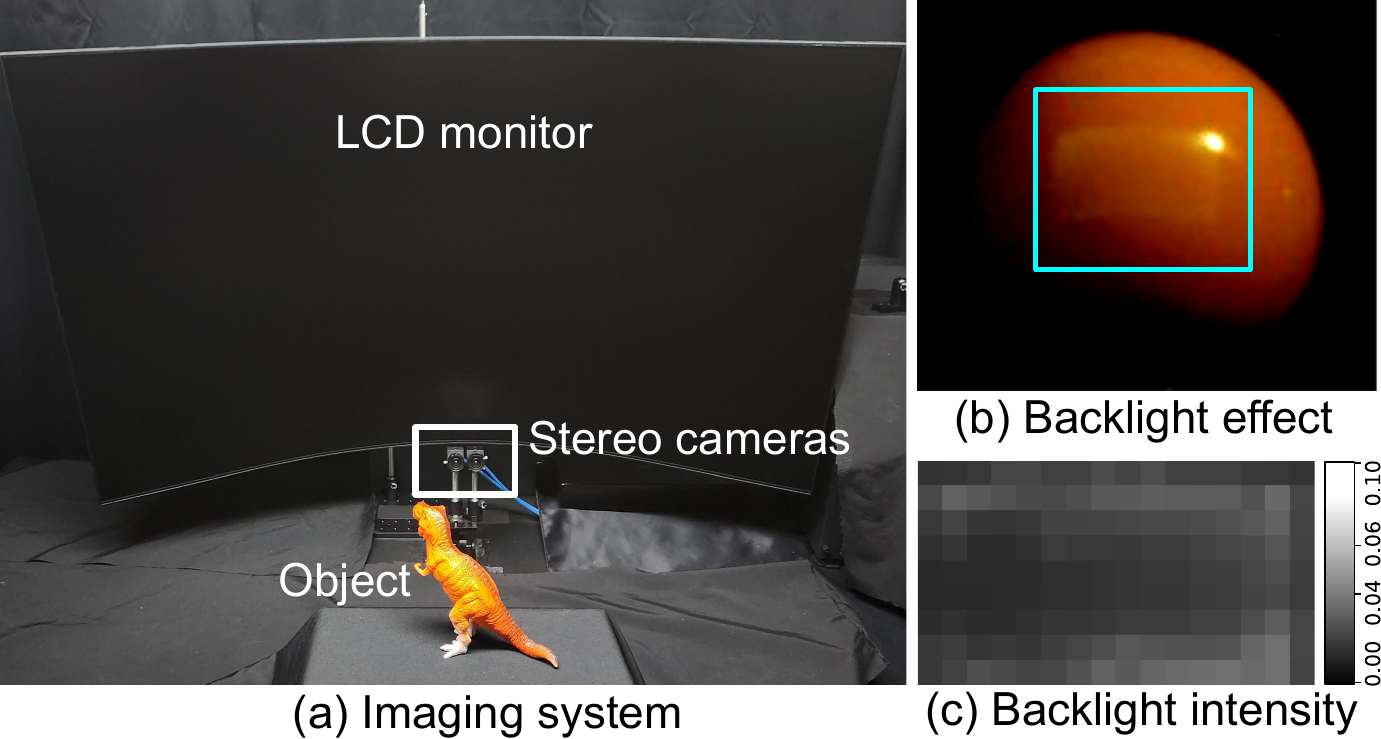}
    \includegraphics[width=\linewidth]{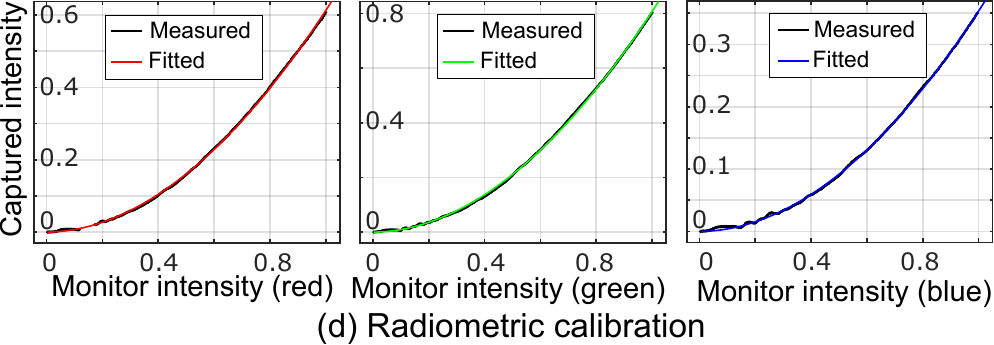}
    \caption{\textbf{Display-camera imaging system.} (a) Our imaging system consists of an LCD monitor and stereo polarization cameras. (b) The LCD monitor exhibits spatially-varying backlight as shown in one of the OLAT images, which (c) we calibrate for accurate inverse rendering. (d) We also obtain the non-linearity of the monitor intensity.}
    \label{fig:system}
    \vspace{-3mm}
\end{figure}

\paragraph{Setup}
To acquire a real-world dataset for display inverse rendering, we built a display-camera system, shown in Figure~\ref{fig:system}(a). Our setup consists of an LCD monitor (Samsung Odyssey Ark) and stereo polarization RGB cameras (FLIR BFS-U3-51S5PC-C) equipped with 8 mm focal-length lenses, covering $30^\circ$ field of view.
The LCD monitor emits vertically polarized light based on the principles of LCD~\cite{heilmeier1968dynamic}. The monitor maximum brightness is 600 $\mathrm{cd}/\mathrm{m}^2$, and each pixel only outputs a maximum intensity of 0.06 $\mathrm{mcd}$, which is too dim to capture even with maximum-exposure imaging.
Following \cite{choi2024differentiable}, we parameterize display pixels using $144=16\times9$ superpixels, where each superpixel consists of $240\times240$ display pixels.
Thus, we represent the display pattern as $\mathcal{L}=\{L_1, \dots, L_N\}$, where each superpixel has an RGB intensity $L_i$, and $N$ denotes the total number of {superpixels}. 
The polarization RGB cameras capture the linearly-polarized light intensity for the R, G, and B channels at $0^\circ$, $45^\circ$, $90^\circ$, and $135^\circ${\cite{baek2021polarimetric}}.

\paragraph{Display Backlight and Nonlinearity}
LCDs often cannot achieve complete darkness even when set to a black value as shown in Figure~\ref{fig:system}(b). 
Modeling this backlight is crucial, as backlight from all display pixels becomes visible in the captured images.
Also, the display intensity is nonlinearly mapped to the value to set, which should be also calibrated.
Taking these into account, we model the $i$-th display superpixel light intensity given the corresponding RGB pattern value we set to display $P_i$ as
\begin{equation}
\label{eq:pattern}
    L_i=s(P_i+B_i)^\gamma,
\end{equation}
where $s$ is a global scalar, $\gamma$ is the non-linear mapping exponent, and $B_i$ is the corresponding spatially-varying backlight intensity.
To calibrate $s$, $B_i$, and $\gamma$, we captured a spherical object with known geometry and reflectance under OLAT patterns, and optimize the three parameters with a loss that minimizes the difference between the OLAT captured images and rendered OLAT images.
Figure~\ref{fig:system}(c) shows the calibrated spatially-varying backlight that resembles the visible backlight in Figure~\ref{fig:system}(b).

\paragraph{Geometric Calibration}
We calibrate the stereo-camera intrinsic and extrinsic parameters using the checkerboard method~\cite{zhang2000flexible}.
We then estimate the position of each display superpixel relative to the reference {left} camera using the mirror-based checkerboard method~\cite{choi2024differentiable}.

\paragraph{Image Formation}
When illuminating a scene point with a display pattern $\mathcal{L}$, the captured intensity by a camera is modeled as:
\begin{align}
\label{eq:image_formation}
I = \mathrm{clip}\left( \sum_{i=1}^{N} (\mathbf{n}\cdot\mathbf{i}) f(\mathbf{i},\mathbf{o}) 
\frac{L_i}{d_i^2}   + \epsilon\right),
\end{align}
where $f$ is the BRDF, $\mathbf{n}$ is the surface normal, $\mathbf{i}$ is the incident light direction from the $i$-th display superpixel, $\mathbf{o}$ is the outgoing view vector, and $d_i$ is the distance from the $i$-th display superpixel to the scene point. The function $\mathrm{clip}(\cdot)$ applies clipping to the camera dynamic range, and $\epsilon$ is Gaussian noise.

\section{Display Inverse Rendering Dataset}
\label{sec:dataset}

\begin{figure*}[h]
    \centering
        \includegraphics[width=0.9\textwidth]{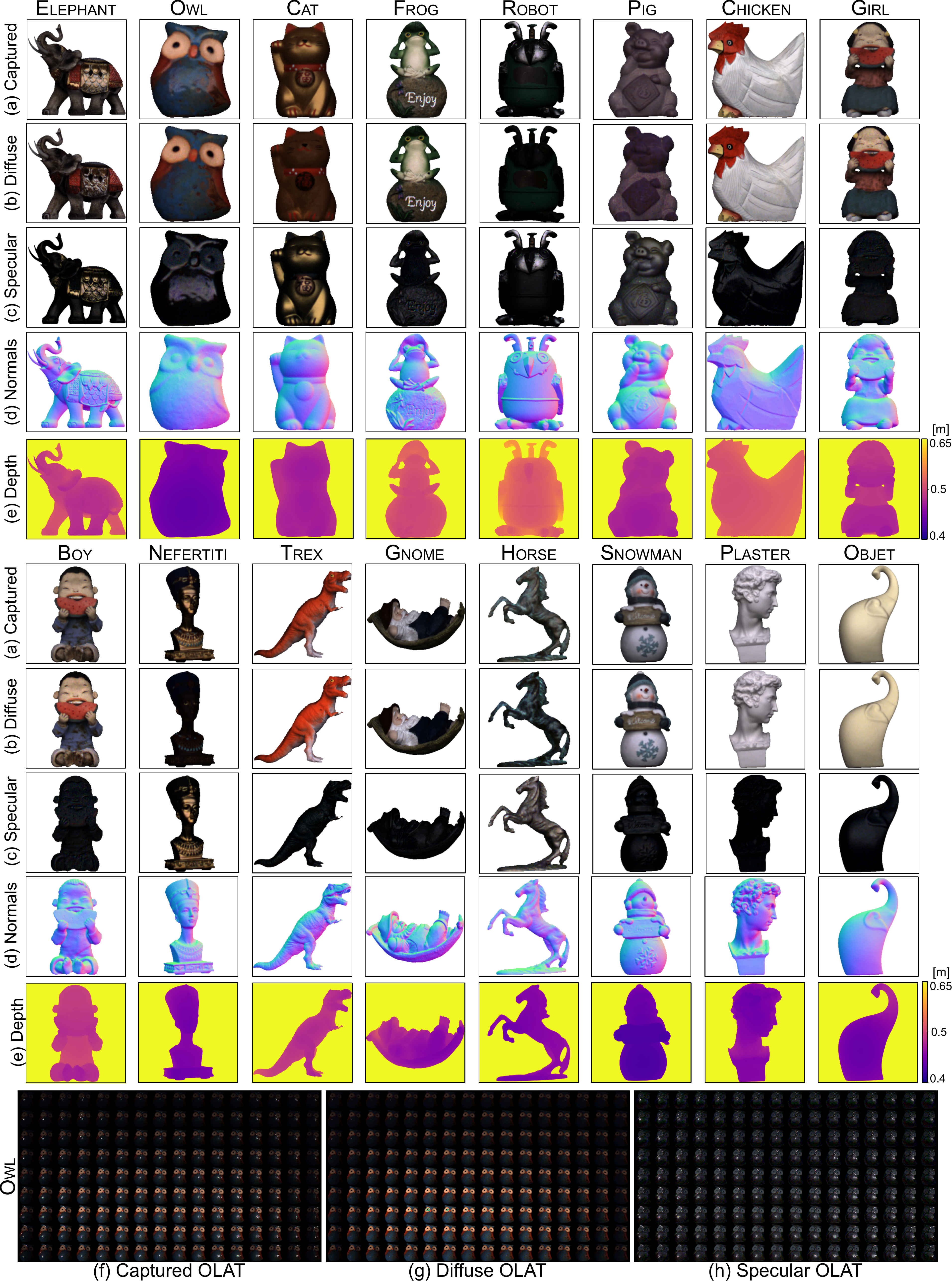}
        \vspace{-3mm}
        \caption{\textbf{Display Inverse Rendering Dataset.} We introduce the first display inverse rendering dataset. 
        We obtain (a) combined, (b) diffuse, and (c) specular stereo images captured under (f–h) OLAT patterns. We provide ground-truth (d) normal maps and (e) depth maps.  }
        \vspace{-3mm}
    \label{fig:dataset}
\end{figure*}

Figure~\ref{fig:dataset} shows our real-world dataset for display inverse rendering. 
Each object has corresponding stereo-polarization RGB images captured under OLAT patterns, ground-truth depth maps, normal maps, and object masks.

\paragraph{Objects}
{We captured 16 objects made of various materials and reflectances from diffuse to specular: resin (\textsc{Frog}, \textsc{Pig}, \textsc{Gnome}, \textsc{Snowman}), ceramic (\textsc{Owl}, \textsc{Objet}), metallic paint (\textsc{Cat}, \textsc{Robot}, \textsc{Nefertiti}), wood (\textsc{Chicken}), clay (\textsc{Girl}, \textsc{Boy}), plastic (\textsc{Trex}), bronze (\textsc{Horse}), plaster (\textsc{Plaster}), and composite (\textsc{Elephant}).
In terms of shape, the objects range from those with simple forms (\textsc{Owl}, \textsc{Cat}, \textsc{Pig}, \textsc{Objet}, \textsc{Chicken}) to those featuring tiny parts (\textsc{Nefertiti}), thin structures (\textsc{Horse}, \textsc{Snowman}), complex details (\textsc{Elephant}, \textsc{Trex}) and curvature (\textsc{Plaster}), as well as concave parts (\textsc{Frog}, \textsc{Girl}, \textsc{Boy}, \textsc{Gnome}, \textsc{Robot}).}
The object sizes range from 8\,cm to 25\,cm.
Objects are placed at 50\,cm from the cameras for the capture.

\paragraph{Ground-truth Geometry}
To obtain ground-truth object shapes, we use structured-light scanning with a high-precision 3D scanner (EinScan SP V2), with a precision tolerance of 0.05 mm. {We align the scanned 3D meshes to the captured images using the mutual information method~\cite{corsini2009image}. Subsequently, we render depth maps, normal maps, and object masks for the camera views on Mitsuba3~\cite{Mitsuba3}.}

\paragraph{Polarimetric Image Processing}
We first convert the captured polarization images at $0^\circ$, $45^\circ$, $90^\circ$, and $135^\circ$ as $\{I_\theta\}_{\theta \in \{0^\circ, 45^\circ, 90^\circ, 135^\circ\}}$ into linear-polarization Stokes-vector RGB images~{\cite{collett2005field}}:
\begin{align}
\label{eq:polar_decomp}
s_0 = \frac{\sum_\theta{I_{\theta}}}{2}, s_1 = I_{0^\circ} - I_{90^\circ},  s_2 = I_{45^\circ} - I_{135^\circ}.
~~\end{align}
Specular reflection tends to maintain the polarization state of display light whereas diffuse reflection becomes mostly unpolarized~\cite{choi2024differentiable}.
This enables us to obtain specular and diffuse images as $I_\text{specular}=\sqrt{(s_1)^2+(s_2)^2}$ and $I_\text{diffuse}=s_0-I_\text{specular}$, respectively, which are shown in Figure~\ref{fig:dataset}.
\clearpage

\begin{figure}[t]
    \includegraphics[width=\linewidth]{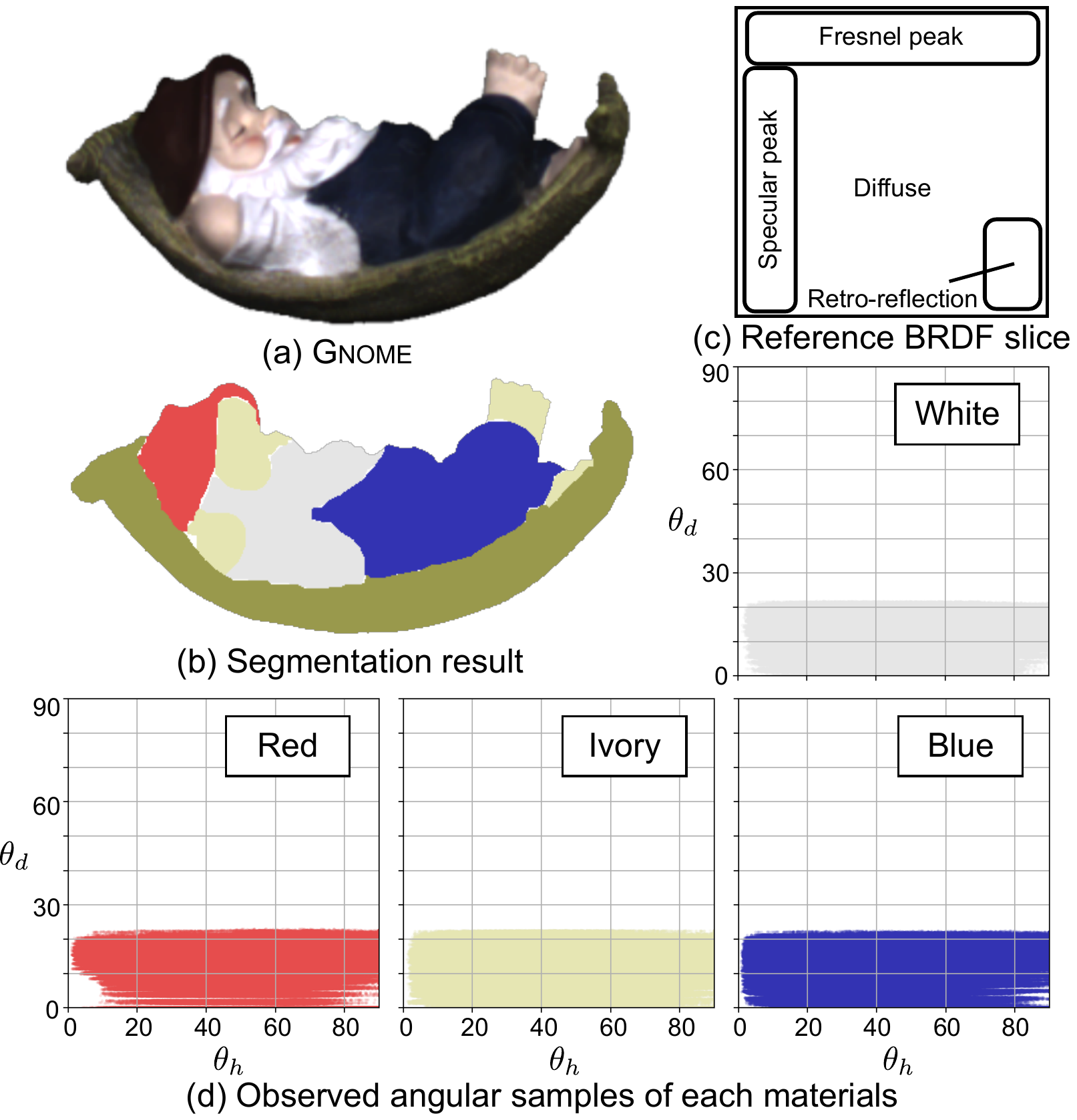}
    \caption{\textbf{Light-view angular samples.} Our display-camera system captures limited light-view angular samples. (a)\&(b) For a segmented scene, (d) we show the sample plots of four segments in $\theta_d, \theta_h$ Rusinkiewicz space~\cite{rusinkiewicz1998new}. (c) The sampled region corresponds to the typical specular, diffuse, and grazing reflections~\cite{nielsen2015optimal}, allowing for inverse rendering.}
    \label{fig:angular_domain}
    \vspace{-5mm}
\end{figure}

\paragraph{Light-view Angular Samples}
Display inverse rendering poses challenges due to the limited coverage of light-view angular samples. In Figure~\ref{fig:angular_domain}, we examine the angular distribution of light-view samples for the segmented four material components. While a full BRDF requires sampling across all Rusinkiewicz coordinates~\cite{rusinkiewicz1998new}, the display-camera setup provides only partial coverage, particularly in terms of $\theta_d$, the angle between the half-way vector and the illumination vector. However, it is worth noting that the half-way angle $\theta_h$ is well-covered from 0 to $\pi/2$, enabling effective sampling of the specular lobe. Additionally, the sampled region corresponds to both diffuse and specular reflections{\cite{nielsen2015optimal}}—a key factor that makes inverse rendering feasible.

\paragraph{Simulation for an Arbitrary Display Pattern}
Leveraging the linearity of incoherent light transport, we simulate a scene illuminated by an arbitrary display pattern $\mathcal{P}=\{P_1,\cdots,P_N\}$, using Equation~\eqref{eq:image_formation} and Equation~\eqref{eq:pattern}, as:
\begin{align}
\label{eq:relighting}
I\left(\mathcal{P}\right) = \mathrm{clip}\left(\sum_{i=1}^{N} I_i s(P_i+B_i)^\gamma   + \epsilon\right),
\end{align}
where $P_i$ is the display superpixel RGB value, $I_i$ is the captured image under the $i$-th OLAT illumination.
The standard deviation of the Gaussian noise $\epsilon$ can be adjusted to reflect different noise levels.

\section{A Baseline for Display Inverse Rendering}
\label{sec:method}
We propose a simple yet effective baseline for display inverse rendering, designed to handle inputs captured under $M$ arbitrary display patterns,  ${\mathcal{P}_1, \cdots, \mathcal{P}_M}$.
As an initialization step, we estimate the normal map using the analytical RGB photometric stereo method~\cite{choi2024differentiable}, which leverages $M$ captured images. Additionally, we estimate a depth map by using the averaged stereo images across multiple patterns as inputs to RAFT stereo~\cite{lipson2021raft}.
Given these normal map and depth map, we optimize the normal map and the reflectance (diffuse albedo, specular albedo, and roughness) of the Cook-Torrance BRDF model. To address the limitations of light-view angular sampling in the display-camera system, we adopt the basis BRDF representation, which models spatially varying BRDFs as a weighted sum of basis BRDFs~\cite{chung2024differentiable, chung2024interpretable, li2022neural}. Specifically, we use the analytic Cook-Torrance model to define each basis BRDF.
We then differentiably render reference-view images for the display patterns ${\mathcal{P}_1, \cdots, \mathcal{P}_M}$ by implementing Equation~\ref{eq:image_formation} in PyTorch and iteratively update the scene representation—comprising normals, basis BRDFs, and their weight maps—by minimizing the RMSE error between the rendered and input images.
Despite challenges such as limited light-view angular samples, display backlight, and near-field lighting in the display-camera setup, our baseline approach enables effective inverse rendering {in only 150 seconds}.

\section{Evaluation}
\label{sec:assessment}
We assess previous photometric stereo methods, inverse rendering approaches, and our proposed baseline method (Section~\ref{sec:method}) using our display-camera dataset.

\begin{table*}[t]
    \centering
    \resizebox{\textwidth}{!}{
    \begin{tabular}{l|cccccccccccccccc}
        \toprule[1pt]
                    &\textsc{Elephant} &\textsc{Owl}   &\textsc{Cat}  &\textsc{Frog}  &{\textsc{Robot}} &{\textsc{Pig}}   &{\textsc{Chicken}}  &\textsc{Girl}  &{\textsc{Boy}} &{\textsc{Nefertiti}}   &{\textsc{Trex}}  &\textsc{Gnome} &{\textsc{Horse}}   &{\textsc{Snowman}}  &\textsc{Plaster} & \textsc{Objet} \\ \hline
        \rowcolor{red!20}  \textbf{Woodham}~\cite{woodham1980photometric}                         &{27.02}  &{26.60} &{21.05} &{21.58} &28.18 &17.02 &18.39 &24.86 & 21.44 &37.03 &18.98 &19.83 &19.27 &32.21 &19.56 &17.28  \\
        \rowcolor{red!20}  \textbf{PS-FCN}~\cite{chen2018ps}                                       &\underline{20.26}  &{15.17} &{10.61} &{19.15} &16.68 &\underline{15.80 }&\underline{11.91} & 25.96 &22.27 &\textbf{20.03} &18.22 &19.33 &\underline{17.48} &18.75 &17.25 &\textbf{7.73} \\
        \rowcolor{red!20}  \textbf{PS-Transformer}~\cite{ikehata2022ps}                            &{26.42}  &{36.43} &{21.11} &{35.34} & 27.31 &49.10 &16.20 &38.66 &35.91 &30.64 &29.86 &36.53 &35.06 &54.26 &33.97 &24.06 \\
        \rowcolor{red!20}  \textbf{SRSH}~\cite{li2022neural}                                       &{26.21}  &{18.49} &{16.95} &{23.42} &19.09 &32.76 &17.88 &37.14 &31.19 &23.97 &25.05 &27.44 &27.70 &27.96 &26.93 &21.87  \\
        \rowcolor{blue!20} \textbf{SCPS-NIR}~\cite{li2022self}                                     &{22.75}  &\textbf{{7.93}} &\textbf{{8.97}} &{16.28} &17.87 &34.89 &\textbf{10.43} &45.12 &37.18 &52.97 &21.85 &16.64 &48.98 &\underline{15.65} &21.30 &\underline{7.94} \\
        \rowcolor{blue!20} \textbf{UniPS}~\cite{ikehata2022universal}                              &25.14 &17.34 &19.69 &24.09 &22.03 &25.77 &22.94 &26.06 &30.00&28.55 &21.64 &24.32 &27.24 &18.86 &19.70 &15.90 \\
        \rowcolor{blue!20} \textbf{UniPS}~\cite{ikehata2022universal} ({$M$}=64)              &{24.93}  &{18.33} &{19.54} &{24.99} &22.18 &25.72 &23.07 &26.38 &30.65 &28.71 &21.86 &24.48 &26.72 &18.89 & 19.43 &16.39 \\
        \rowcolor{blue!20} \textbf{SDM-UniPS}~\cite{ikehata2023scalable} ({$M$}=64)           &\textbf{18.83} &14.37 &9.70 &\textbf{14.12}&\textbf{14.85}&\textbf{15.33}&16.05 &\textbf{14.99}&\textbf{15.22}&\underline{22.73} &\textbf{14.58}&\textbf{13.46}&\textbf{16.93}&\textbf{15.18 }&\textbf{12.55}&{9.38} \\
        \rowcolor{blue!20} \textbf{SDM-UniPS}~\cite{ikehata2023scalable} ({$M$}=10)           &{20.53}  &\underline{12.77} &\underline{9.43} &\underline{15.23} &\underline{16.48} &{16.12} &16.10 &\underline{15.23} &\underline{17.25} &24.32 &\underline{15.36} &\underline{15.47} &17.62 &16.57 &\underline{13.39} &9.58 \\
        \bottomrule[1pt]  
        \end{tabular}}
        \caption{\textbf{Photometric-stereo evaluation using OLAT patterns.} Normal reconstruction error in Mean Angular Error (MAE) for calibrated (red) and uncalibrated (blue) photometric stereo. Highest performance in \textbf{bold} and the second-best in \underline{underline}. When $M$ is specified, it means the $M$ number of uniform-sampled OLAT patterns is used for evaluation. 
        }
        \vspace{-4mm}
    \label{tab:photometric_stereo}
\end{table*}

\begin{table}[t]
    \centering
    \resizebox{0.9\linewidth}{!}{
    \begin{tabular}{c|cccccc}
        \toprule[1pt]
        
                        Method        & {\textbf{Ours}} &{\textbf{Ours}}   & \textbf{SRSH}~\cite{li2022neural}             & \textbf{DPIR}~\cite{chung2024differentiable}                         & \textbf{IIR}~\cite{chung2024interpretable}                     \\ 
                        Patterns        &                Multiplexed                   & OLAT             &   OLAT    &  OLAT    & OLAT \\ \hline
            \textbf{PSNR} \textbf{[dB] $\uparrow$}                   &{{37.27}}    &{\underline{39.33}}  &\textbf{{41.28}}  & {{34.30}} & {{38.20}}\\
            \textbf{SSIM}{$\uparrow$}                      &{{0.9766}}                &{{0.9821}}   &\textbf{0.9895} & {{0.9790}} & {\underline{0.9850}} \\
            \hline
            \textbf{MAE}  \textbf{[${\cdot}^{\circ}$] $\downarrow$} &{\underline{23.97}} &{\textbf{20.94}}&{25.25} &{41.09} &{38.38} \\  
        \bottomrule[1pt]  
        \end{tabular}}
        \caption{\textbf{Inverse-rendering evaluation.} Our baseline method enables high-quality relighting accuracy in PSNR and SSIM (first two rows) and normal accuracy in MAE (last row) for both OLAT and multiplexed patterns. While SRSH enables effective relighting, the normal accuracy is low and non-trivial to support multiplexed patterns. 
        }
        \vspace{-4mm}
    \label{tab:inverse_rendering}
\end{table}

\begin{table}
    \centering
    \resizebox{\linewidth}{!}{
    \begin{tabular}{l|ccc|ccc}
        \toprule[1pt]
        &\multicolumn{3}{c}{Learned display pattern~\cite{choi2024differentiable} $\downarrow$}      &\multicolumn{3}{c}{Heuristic display pattern $\downarrow$}                         \\ \cline{2-7}
        &{$M$=2}  &{$M$=4}  &{$M$=10}                   &{(a) $M$=2}  &{(b) $M$=4}  &{(c) $M$=10}  \\ \hline
        UniPS~\cite{ikehata2022universal}                           &27.7078    &25.9408    &\underline{25.7541}                   &65.7171    &63.1694    &{63.4573}                   \\
        SDM-UniPS~\cite{ikehata2023scalable}                        &\textbf{23.5079}    &\textbf{19.8946}    &\textbf{18.1829} &\underline{42.3576}    &\textbf{29.9320}    &\textbf{32.0718} \\
        DDPS~\cite{choi2024differentiable}     &\underline{24.5678}    &\underline{23.3800}    &29.3716                           &\textbf{32.0480}    &\underline{35.1451}    &\underline{36.5606      }                     \\    
        \bottomrule[1pt]  
        \end{tabular}}
        \caption{\textbf{Multiplexed patterns with varying numbers.} 
        We evaluate normal reconstruction accuracy of photometric stereo methods using various numbers of heuristic patterns and learned display patterns.
        }
        \vspace{-3mm}
    \label{tab:pattern_photometric_stereo}
\end{table}

\begin{table}
    \centering
    \resizebox{0.85\linewidth}{!}{
    \begin{tabular}{l|ccc}
        \toprule[1pt]
        &\multicolumn{3}{c}{Learned display pattern~\cite{choi2024differentiable} $\downarrow$}                               \\ \cline{2-4}
        &{$M$=2}  &{$M$=4}  &{$M$=10}                    \\ \hline
        DDPS~\cite{choi2024differentiable}  (Diffuse +Specular)     &{24.5678}          &{23.3800}           &29.3716                                        \\    
        DDPS~\cite{choi2024differentiable} (Diffuse)                &\textbf{23.2807}   &\underline{21.2126}    &\underline{27.7281}         \\    
        SDM-UniPS~\cite{ikehata2023scalable}   (Diffuse +Specular))         &\underline{23.5079}    &\textbf{19.8946}    &\textbf{18.1829}  \\
        SDM-UniPS~\cite{ikehata2023scalable} (Diffuse)              &{35.0658}          &{31.2040}           &{30.1058}  \\
        \bottomrule[1pt]  
        \end{tabular}}
        \caption{\textbf{Photometric stereo with diffuse component and varying number of patterns.} 
        We evaluate the impact of using diffuse images rather than the captured one containing both diffuse and specular components.
        }
        \vspace{-5mm}
    \label{tab:diffuse_photometric_stereo}
\end{table}

\begin{figure}[t]
    \includegraphics[width=\linewidth]{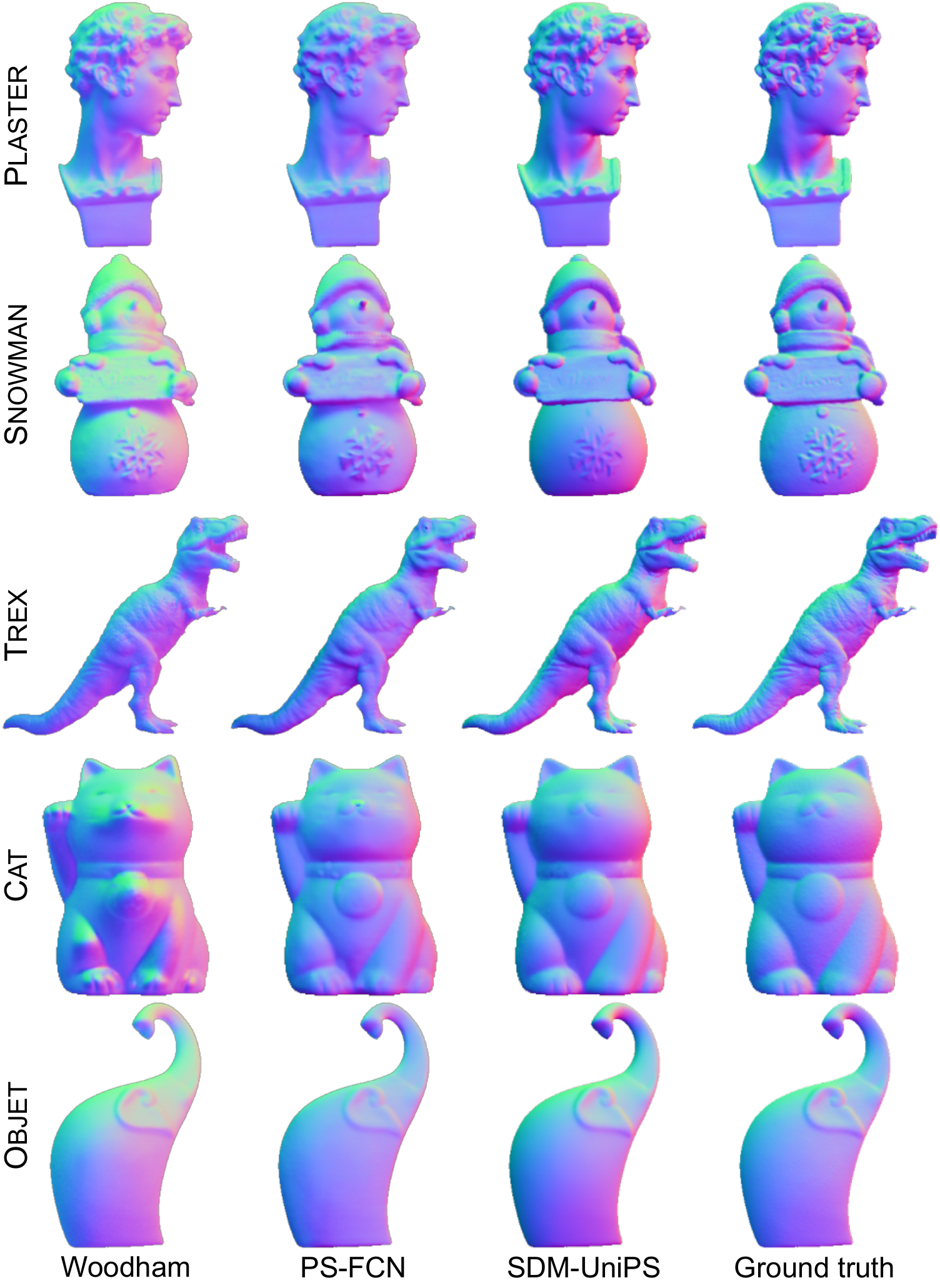}
    \caption{\textbf{Photometric stereo with OLAT patterns.}
    SDM-UniPS~\cite{ikehata2023scalable} demonstrates highly accurate normal reconstruction results, outperforming other methods.}
    \label{fig:ps}
    \vspace{-5mm}
\end{figure}

\paragraph{Photometric Stereo using OLAT Patterns}
Photometric stereo is a subtask of inverse rendering that focuses on normal reconstruction. We evaluate both calibrated~\cite{woodham1980photometric, chen2018ps, ikehata2022ps, li2022neural} and uncalibrated~\cite{li2022self, ikehata2022universal, ikehata2023scalable} methods on our dataset.
As shown in Table~\ref{tab:photometric_stereo} and Figure~\ref{fig:ps}, recent uncalibrated photometric stereo techniques—particularly SDM-UniPS~\cite{ikehata2023scalable}—demonstrate highly accurate normal estimation. 
This indicates that the 144 OLAT images in our display setup provide sufficient information for precise normal reconstruction.

\begin{figure*}[t]
    \includegraphics[width=\textwidth]{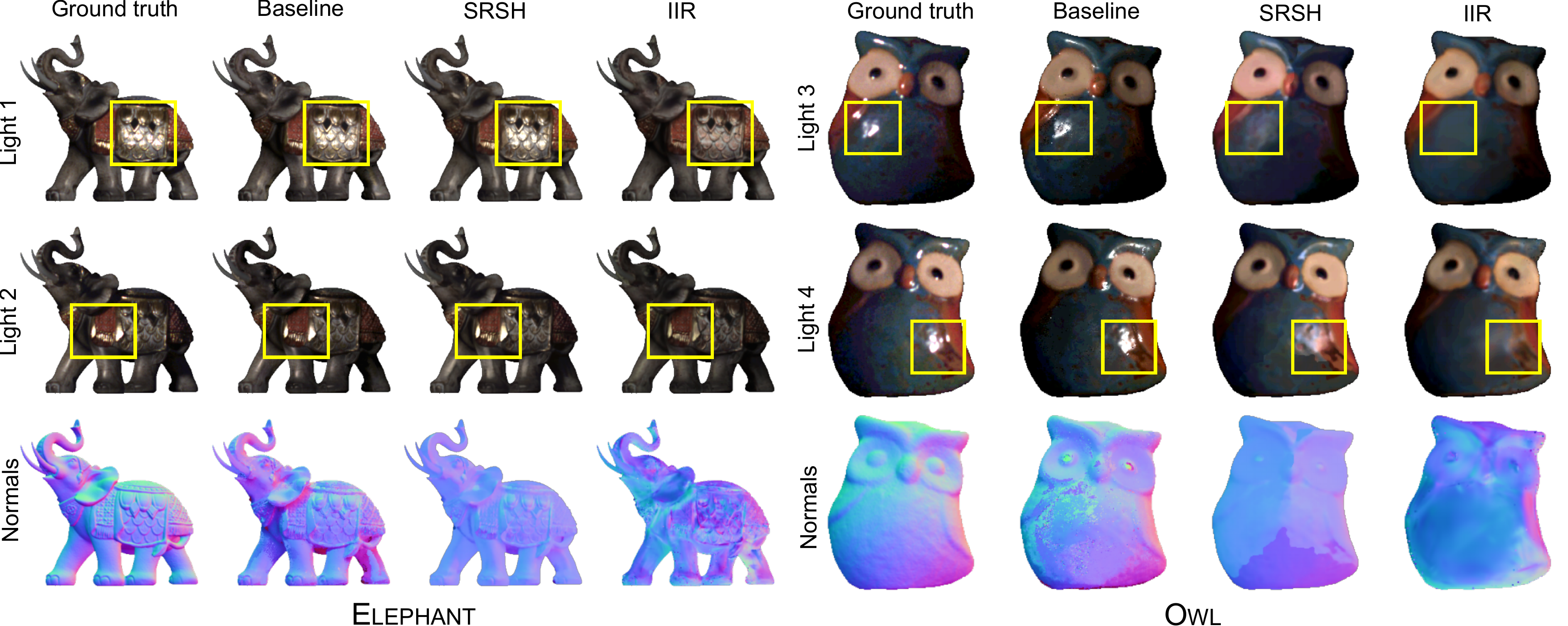}
    \caption{\textbf{Inverse rendering with OLAT patterns.} 
    Our proposed baseline method (second column) achieves qualitatively more accurate relighting and normal reconstruction, outperforming other inverse rendering methods.}
    \vspace{-3mm}
    \label{fig:ir}
\end{figure*}

\paragraph{Inverse Rendering using OLAT Patterns}
Many existing inverse rendering methods cannot be directly applied to the display inverse rendering configuration due to {the inherent challenges such as limited light-view angular samples, backlight, and near-field effects}. 
To evaluate performance in this setting, we test four available inverse rendering methods: one single-view approach~\cite{li2022neural}, two multi-view methods~\cite{chung2024differentiable, chung2024interpretable}, and our proposed baseline model.
For evaluation, we divide the 144 OLAT images into training and testing sets with a 5:1 ratio. As shown in Table~\ref{tab:inverse_rendering} and Figure~\ref{fig:ir}, our proposed baseline model achieves accurate relighting of specular appearances, whereas other methods produce blurry relighting results. 
This demonstrates that our approach effectively handles the challenges of limited light-view angular samples, backlight, and near-field effects, leading to robust display inverse rendering.

\begin{figure}[t]
    \includegraphics[width=\linewidth]{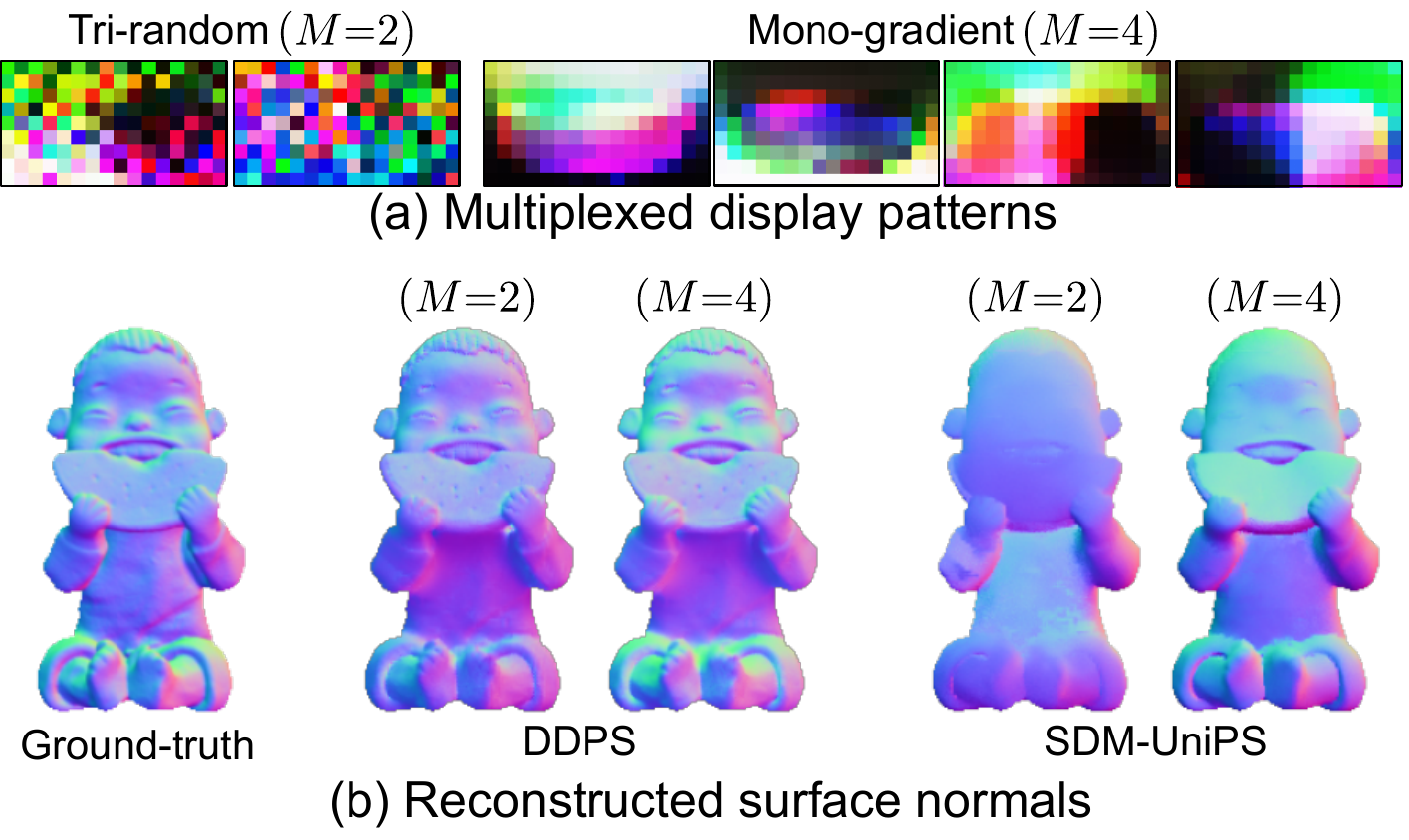}
    \caption{\textbf{Multiplexed display patterns for photometric stereo.}
    We found that analytical photometric stereo such as DDPS~\cite{choi2024differentiable} is more robust to small number of display patterns than the learning-based photometric stereo such as SDM-UniPS.
    }
    \label{fig:display_patterns_ps}
    \vspace{-3mm}
\end{figure}

\begin{figure*}[t]
    \includegraphics[width=\linewidth]{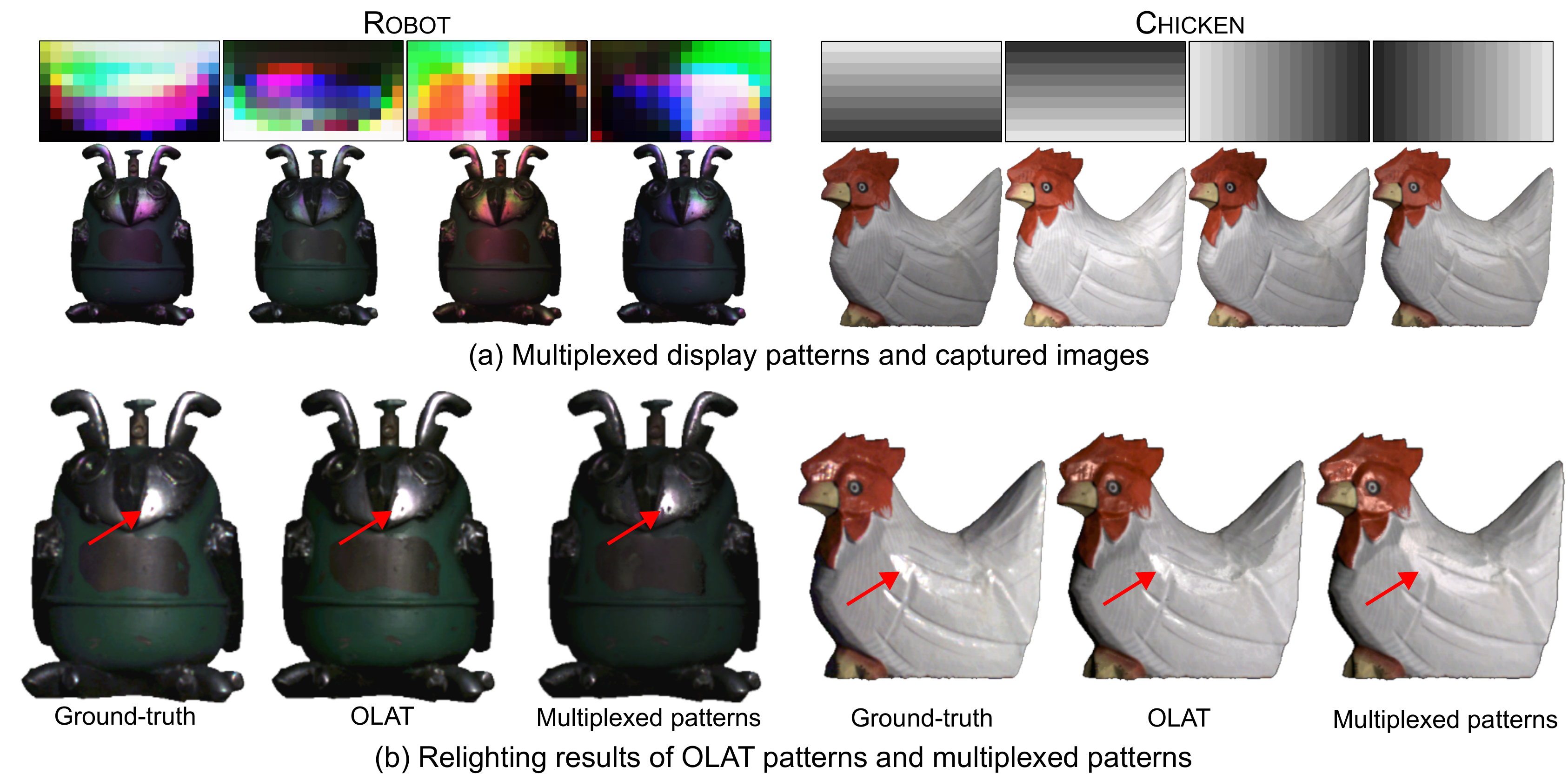}
    \caption{\textbf{Multiplexed display patterns for inverse rendering.}  
    Inverse rendering performed with 144 OLAT patterns achieves relighting results that closely approximate the ground truth. Although inverse rendering can be performed using only four heuristic or learned patterns~\cite{choi2024differentiable}, relighting accuracy remains less accurate than that achieved with OLAT patterns.
    }
    \vspace{-3mm}
    \label{fig:display_patterns_ir}
\end{figure*}

\paragraph{Multiplexed Display Patterns for Photometric Stereo}
While OLAT images provide sufficient information for inverse rendering, capturing all 144 OLAT patterns is time-consuming. A more efficient approach in display-camera systems is to use $M$ multiplexed display patterns, formed as linear combinations of the OLAT patterns.
We evaluated two multiplexed display pattern strategies: {manually-designed} and computationally learned patterns from DDPS~\cite{choi2024differentiable}. As shown in Table~\ref{tab:pattern_photometric_stereo} and Figure~\ref{fig:display_patterns_ps}, even with just two multiplexed patterns, accurate normal reconstruction is achievable.
Additionally, Table~\ref{tab:pattern_photometric_stereo} presents results for the learned “Tri-random ($M$=2)”{~\cite{choi2024differentiable}} and “Mono-gradient ($M$=4)”{~\cite{ma2007rapid}} patterns from DDPS, along with a concatenated pattern ($M$=10) that integrates these with the “Mono-complementary” pattern{~\cite{kampouris2018diffuse}}. For heuristic patterns, we tested the “Tri-complementary ($M$=2)”{~\cite{lattas2022practical}} and “Mono-gradient ($M$=4)” patterns{~\cite{ma2007rapid}}, as well as a concatenated ($M$=10) pattern combining them with the “Mono-complementary” pattern{~\cite{kampouris2018diffuse}}. Our results indicate that learned patterns consistently outperform heuristic patterns when using the same number of patterns. However, simply increasing the number of learned patterns does not always lead to further improvements in performance.
\vspace{-3mm}
\paragraph{Multiplexed Display Patterns for Inverse Rendering}
We evaluate the impact of multiplexed display patterns on our proposed baseline method for inverse rendering.
Table~\ref{tab:inverse_rendering} shows the quantitative results and
Figure~\ref{fig:display_patterns_ir} presents the inverse rendering results using two patterns, each consisting of four images: a monochromatic gradient pattern{~\cite{ma2007rapid}} and a learned display pattern{~\cite{choi2024differentiable}}. {While the relighting results do not achieve the same accuracy as OLAT's results, they still exhibit reasonable performance,} with a relighting PSNR of {38.07 and 37.77} dB respectively.
These findings suggest that designing display patterns that enable efficient capture while enhancing inverse rendering performance remains an open research challenge.

\begin{table}
    \centering
    \resizebox{0.8\linewidth}{!}{
    \begin{tabular}{l|ccc}
        \toprule[1pt]
                                                &\textbf{low res.}  &\textbf{32-inch}  &\multirow{2}{*}{\textbf{Default}}  \\ 
                                                &{($M$=32)}  &{($M$=50)}  &  \\ \hline
        Woodham~\cite{woodham1980photometric}  &55.973    &\underline{29.175}    &{23.144}  \\
        PS-FCN~\cite{chen2018ps}                &\underline{44.516}    &{40.327}    &\underline{17.286}  \\    
        SDM-UniPS~\cite{ikehata2023scalable}    &{\textbf{14.838}}    &{\textbf{15.716}}    &{\textbf{14.896}} \\
        \bottomrule[1pt]  
        \end{tabular}}
        \caption{\textbf{Impact of display configuration.} 
        We found that normal reconstruction error (MAE) of SDM-UniPS~\cite{ikehata2023scalable} is low for different display configurations: our original display setup, low-resolution superpixels, and a 32-inch display size. 
        }
        \vspace{-5mm}
    \label{tab:hardware_configuration}
\end{table}

\paragraph{Impact of using Diffuse Images}
We evaluate the effect of incorporating polarization-separated diffuse images under the same set of display patterns in Table~\ref{tab:diffuse_photometric_stereo}.
As shown in Table~\ref{tab:diffuse_photometric_stereo}, using diffuse images can improve normal reconstruction accuracy and efficiency in capture by reducing the number of required input images. However, this improvement is not consistent across all methods, suggesting that developing reconstruction methods that better use optically-separated diffuse and specular images is a future direction.

\vspace{-4mm}

\paragraph{Impact of Display Specifications}
We evaluate how different display specifications impact inverse rendering performance.
Table~\ref{tab:hardware_configuration} summarizes normal reconstruction results under various conditions, including lower-resolution superpixels and a simulated 32-inch monitor.
When using superpixels smaller than 240×240 pixels to enhance resolution, the captured images remain too dark even at maximum camera exposure, and this is unsuitable for inverse rendering.
Conversely, with 480×480-pixel superpixels arranged in an 8×4 resolution, the display behaves like an area light source, causing both the conventional method~\cite{woodham1980photometric} and PS-FCN methods to fail in normal reconstruction.
However, SDM-UniPS, which accounts for this type of lighting model, maintains relatively stable performance, with errors comparable to those observed when using 32 patterns.
Additionally, when sampling only 10×5 superpixels—corresponding to the physical area of a 32-inch display—the Woodham's method exhibits predictable performance degradation due to a reduced range of incident light angles, while PS-FCN fails to provide reliable estimates under this configuration.
A notable observation in inverse rendering is the impact of removing distant light sources. 
In a 32-inch display setting, these sources are removed and improves the surface normal MAE of SRSH~\cite{li2022neural} from 25.25 to 17.68, highlighting the significant role of light attenuation in display-based setups.
Furthermore, when the baseline model does not account for light attenuation, the PSNR drops from 39.78 to 37.43, confirming the importance of modeling near-field effects.

\section{Conclusion}
\label{sec:conclusion}
In this paper, we introduced the first real-world dataset for display inverse rendering. To construct this dataset, we developed a display-camera imaging system and carefully calibrated the display and camera parameters relevant to inverse rendering.
Using our dataset, we conducted a comprehensive evaluation of existing photometric stereo and inverse rendering methods within the display-camera configuration. Our analysis revealed that current methods require further advancements, particularly in adapting to diverse display patterns, achieving robust reflectance reconstruction under limited light-view angular samples, and leveraging polarization properties inherent to display-camera setups.
We hope that our dataset will serve as a resource, driving future developments and evaluations of inverse rendering methods for display-camera systems.

\paragraph{Future Directions}
Future work could explore advanced methods for effectively exploiting separated diffuse-specular components, as well as methods to handle the challenges posed by limited light-view angular samples. In addition, investigating optimized multiplexed display patterns and their corresponding reconstruction methods presents a promising avenue for further research.
We believe that the dataset we have proposed will serve as a valuable resource, accelerating developments in these area.

\paragraph{Acknowledgments}
Seung-Hwan Baek was partly supported by Korea NRF grants (RS-2023-00211658, RS-2024-00438532), an IITP-ITRC grant (RS-2024-00437866), and a KEIT grant (RS-2024-0045788), funded by the Korea government (MSIT, MOTIE).

\clearpage

{
    \small
    \bibliographystyle{ieeenat_fullname}
    \bibliography{references}
}
\end{document}


\begin{CJK}{UTF8}{}
\CJKfamily{mj}

\title{A Real-world Display Inverse Rendering Dataset

-Supplemental Document-}

\author{Seokjun Choi\footnotemark[1] ~ ~ ~
Hoon-Gyu Chung\footnotemark[1] ~ ~ ~
Yujin Jeon\footnotemark[1] ~ ~ ~
Giljoo Nam\footnotemark[2] ~ ~ ~
Seung-Hwan Baek\footnotemark[1] \\
\footnotemark[1]~ POSTECH   ~ ~ ~   ~ ~ ~  \footnotemark[2]~ Meta\\
}

\maketitle

In this supplemental document, we provide additional results and details in support of our findings in the main manuscript. 

\tableofcontents

\section{Details on Imaging System}
\paragraph{Lighting Module}
For our display module, we employ a commercially available large curved LCD monitor (Samsung Odyssey Ark). This display features a 55-inch liquid-crystal panel with a resolution of 2160$\times$3840 pixels and a peak brightness of 600 $cd/m^2$. Owing to the polarization-sensitive optical components inherent in LCD technology, each pixel emits horizontally linearly-polarized light spanning the trichromatic RGB spectrum. 

\paragraph{Capture Module}
We further utilize two polarization camera (FLIR BFS-U3-51S5PC-C) that integrates on-sensor linear polarization filters oriented at four distinct angles. Consequently, the camera records four polarized light intensities at $0^\circ$, $45^\circ$, $90^\circ$, and $135^\circ$, denoted as $I_{0^\circ}$, $I_{45^\circ}$, $I_{90^\circ}$, and $I_{135^\circ}$, respectively.
The sensor captures a linear raw signal, to which we apply a series of linear image processing steps, including black level subtraction, demosaicing, and undistortion.
All images in the dataset were captured using a fixed shutter time of 440\,ms under a single-exposure setting.
{As a cost-effective alternative to a high-end polarization camera, one may adopt a conventional camera augmented with a linear-polarization film. Aligning the film’s polarization axis perpendicular to that of the display facilitates the capture of diffuse image components.}

\paragraph{Device Control}
To manage the display outputs and control the polarization camera, we employ the PyGame and PySpin libraries, respectively. The devices are interfaced with a desktop computer via an HDMI cable and a USB3 connection, with software synchronization ensuring coordinated operation between the display and the camera. 

\paragraph{Radiometric Calibration}
The monitor’s emitted radiance does not exhibit a linear correlation with the pixel values of the display pattern. To correct for this nonlinearity, we capture images of gray patches on a color checker across various intensity levels. An exponential function is then fitted to the measured intensities as a function of the monitor's pixel values for each individual color channel.

\begin{figure}[h]
	\centering
		\includegraphics[width=0.8\linewidth]{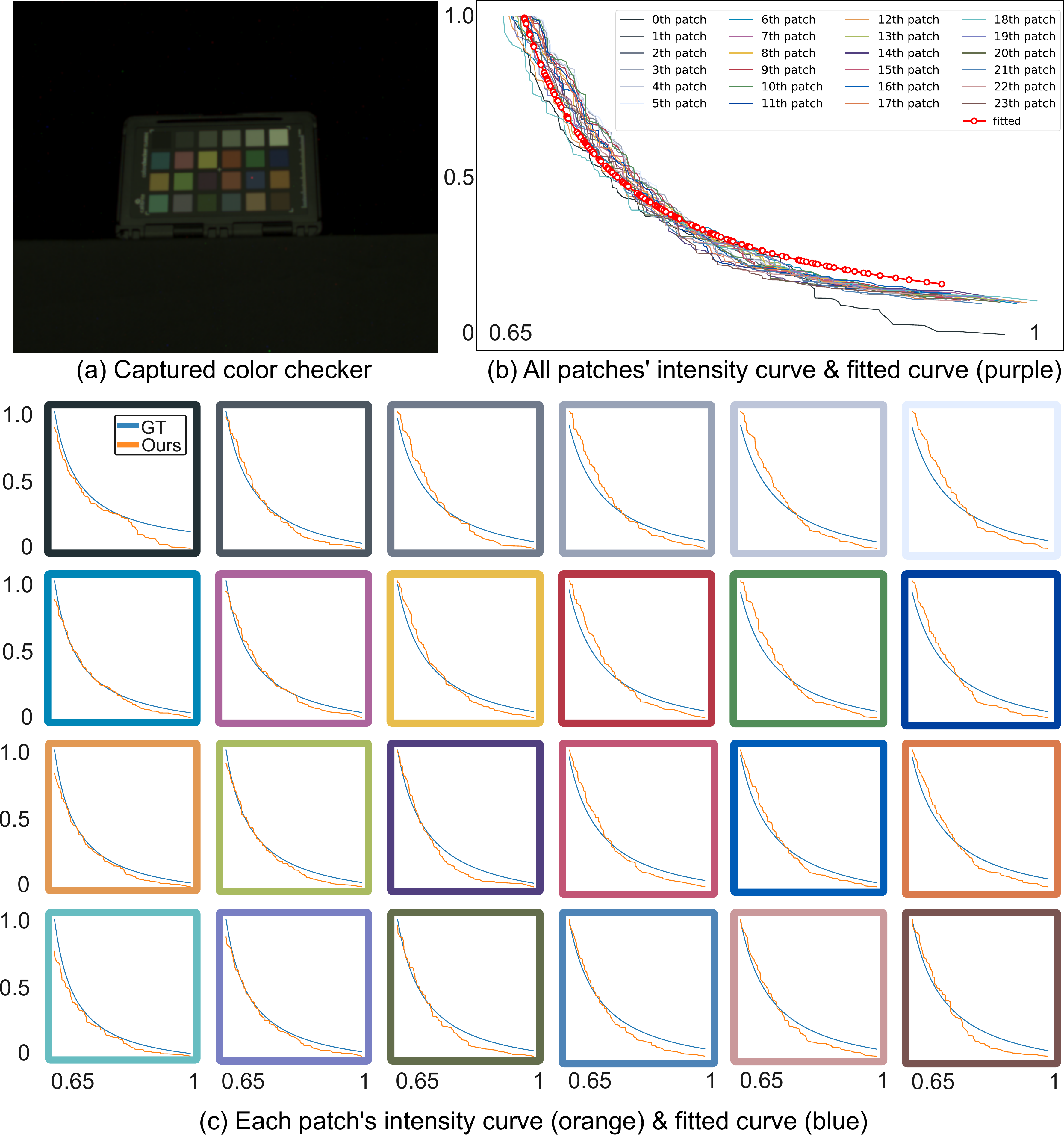}
		\caption{\textbf{Light fall-off calibration.} The parameters are fitted by analyzing the variation in pixel intensity as a function of the distance to the light source. 
        }
  		\label{fig:attenuation}
\end{figure}

\paragraph{Light attenuation calibration}
To calibrate the light attenuation effect as a function of the distance between the object and the light source, a color checker was positioned in front of the imaging system and an OLAT image was captured as shown in Figure~\ref{fig:attenuation}.(a). For each light position, the brightness variation of the color patches is measured, and the parameters $a$, $b$, and $c$ in the following equation are fitted:
\begin{equation}    
\frac{1}{{a+b\times{distance}^2}} +c
\end{equation}

Figure~\ref{fig:attenuation}. (b) and (c) shows curves fitted to captured intensity variation.

\begin{figure}[h]
	\centering
		\includegraphics[width=0.55\linewidth]{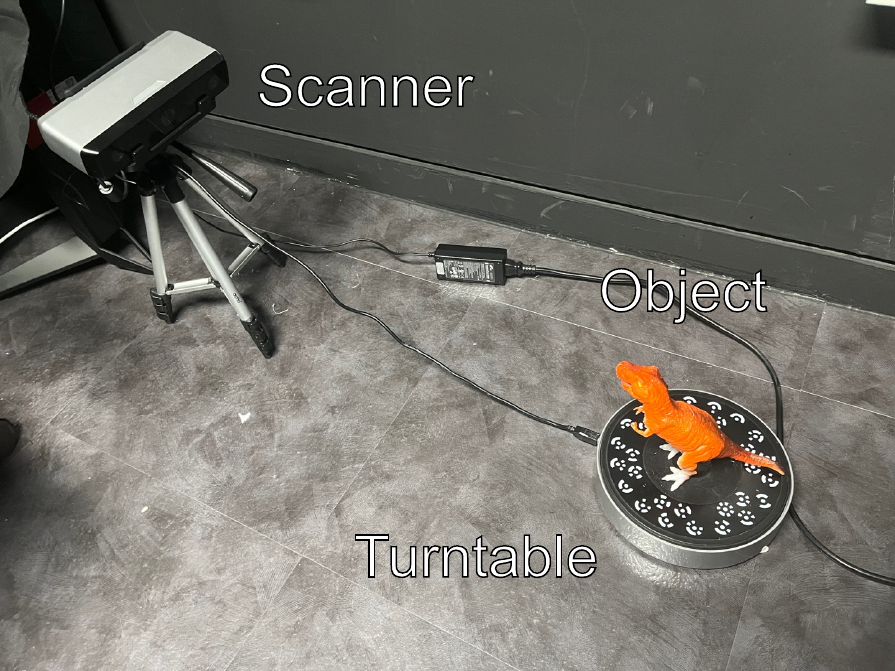}
		\caption{\textbf{EinScan SP V2.} An object is placed on turntable, and the scanner is looking at the object.}
  		\label{fig:scanner}
\end{figure}

\section{Details on Scan System}
We employed the EinScan SP V2 scanner to acquire mesh data of three-dimensional objects. The EinScan SP V2 is a structured light active stereo imaging device that scans objects placed on a turntable. Device calibration is performed by positioning a checkerboard on the turntable. During a single rotation, eight images are captured to generate data points, which are then registered to reconstruct the object's geometry. The system maintains an accuracy within a tolerance of 0.05\,mm.
The scan scenario is shown in Figure~\ref{fig:scanner}.

\paragraph{Mesh-raster alignment}
The scanned mesh is aligned with the captured images to extract the ground truth depth map, normal map, and mask. The registration between the images and the mesh is performed in a semi-automatic and semi-manual manner using the mutual information methods~\cite{corsini2009image} available in MeshLab. The mesh pose obtained through this alignment is then imported into the differentiable renderer, Mitsuba3, to render the depth map, normal map, and mask.

\begin{table*}[t]
    \centering
    \resizebox{\textwidth}{!}{%
        \begin{tabular}{lccccccc}
        \toprule[1pt]
            \multirow{2}{*}{Dataset}                         & Ground-truth & Imaging                & Lighting              & Number of   & Number of    & Number of  \\
                                                             & geometry    & system                 & calibration           & objects     & views        & lights     \\ \hline\hline
            Blobby and Sculpture~\cite{chen2018ps}           & \cmark      & Synthetic              & Direction          &       18&  1           & 64         \\
            CyclePS~\cite{ikehata2018cnn}                    & \cmark      & Synthetic              & Direction          &       45    &     1        & 1300       \\
            PS-Wild~\cite{ikehata2022universal}              & \cmark      & Synthetic              & \xmark             &       410   &     1        & 10         \\ \hline
            Gourd and Apple~\cite{alldrin2008photometric}    & \xmark      & Light rig              & Direction             & 3           & 1            & 112    \\
            Harvard~\cite{xiong2014shading}                  & \xmark      & Light rig              & Direction             &    7        &    1         & 20         \\
            DTU~\cite{jensen2014large}                       & \cmark      & Robot                  & \xmark                &         80  &      64 & 7          \\ 
            MIT-intrinsic~\cite{grosse2009ground}            & \xmark      & Commodity camera       & \xmark                &         20  &      1       & 10         \\
            NeROIC~\cite{kuang2022neroic}                    & \xmark      & Commodity camera       & Env. map              &          3  &       40     & 6      \\
            NeRF-OSR~\cite{rudnev2022nerf}                   & \xmark      & Commodity camera       & Env. map              &         8   & 3240    & 110    \\
            Stanford ORB~\cite{kuang2023stanfordorb}         & \cmark      & Commodity camera       & Env. map              &       14    &       70     & 7          \\
            ReNe~\cite{toschi2023relight}                    & \xmark      & Robot                  & Position              &         20  &      50      & 40         \\
            Light Stage Data Gallery~\cite{chabert2006relighting}& \xmark  & Light stage            & Direction             &  9          & 1            & 253        \\
            Open Illumination~\cite{liu2024openillumination} & \xmark      & Light stage            & Position              &       64    &    72        & 154        \\
            Polar-lightstage~\cite{yang2024acquisition}      & Pseudo      & Light stage            & Direction             &        26   &     8        & 346        \\
            LUCES~\cite{mecca2021luces}                      & \cmark      & Light rig              & Position              &     14      &       1      & 52         \\ 
            DiLiGenT~\cite{shi2016benchmark}                 & \cmark      & Light rig              & Position              &   10        &   1          & 96         \\
            DiLiGenT102~\cite{ren2022diligent102}           & \cmark      & Gantry                 & Direction             & 100         &   1          & 100        \\
            DiLiGenT-PI~\cite{wang2023diligent}             & \cmark      & Gantry                 & Direction             &    34       &   1          & 100        \\
            DiLiGenRT~\cite{guo2024diligenrt}               & \cmark      & Gantry                 & Direction             & 54          &     1        & 100        \\ 
            DiLiGenTMV~\cite{Li2020diligentmv}               & \cmark      & Studio/desktop scanner & Direction             &    5        &       20     & 96         \\ \hline
            \textbf{Ours}                                    & \cmark      & \textbf{Display and camera}     & \textbf{Position}              & \textbf{16}          & \textbf{2}            & \textbf{144}        \\
        \bottomrule[1pt] 
        \end{tabular}
    }
    \caption{\textbf{Inverse rendering datasets.} We classified the datasets related to inverse rendering into their respective categories and organized them into a table.}
    \label{tab:datasets}
\end{table*}

\begin{figure*}[h]
	\centering
		\includegraphics[width=0.8\textwidth]{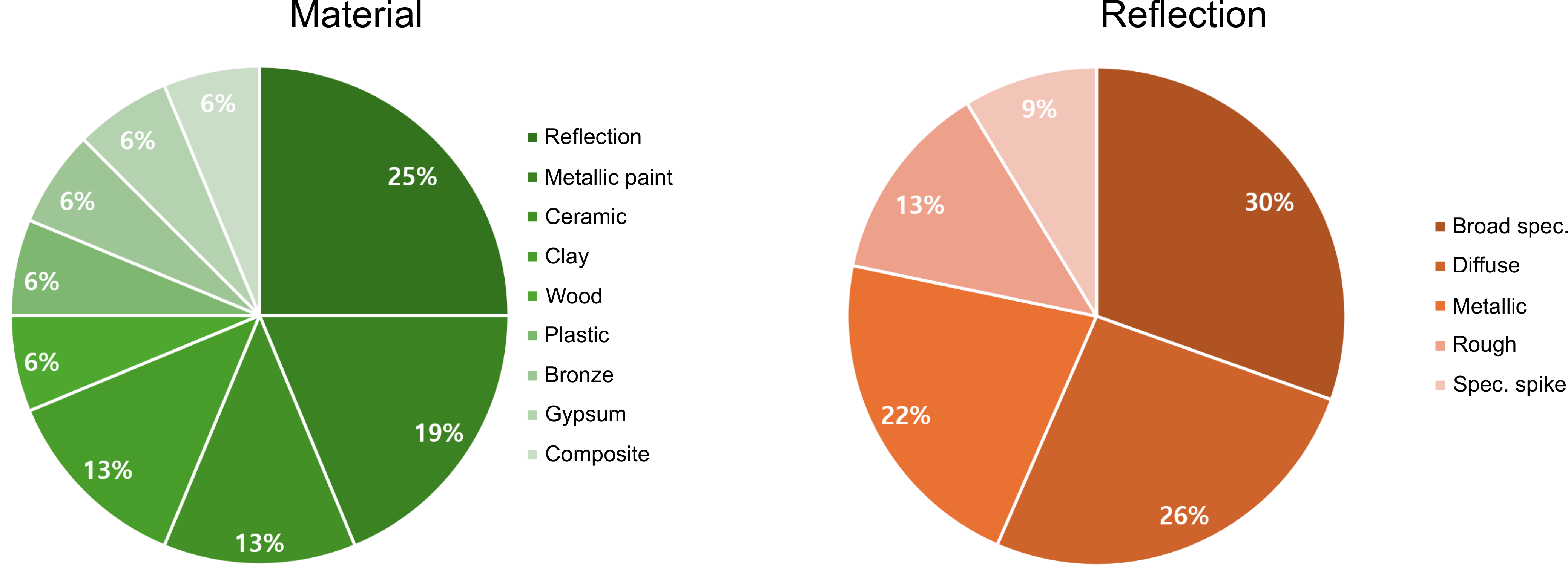}
		\caption{\textbf{Data statistics.} We conducted a statistical analysis of 16 objects based on two aspects: their material properties and surface reflectance characteristics.
}
  		\label{fig:statistics}
\end{figure*}

\section{Dataset}
\paragraph{Comparison with Existing datasets.}
Table~\ref{tab:datasets} presents the existing inverse rendering datasets. Some of these datasets were originally developed for photometric stereo, yet they have also been widely adopted for inverse rendering. Each dataset exhibits unique characteristics based on whether ground truth geometry is provided, the type of illumination and imaging system used, the available lighting calibration information, the number of captured objects, the number of views, and the number of light sources employed. These factors determine the suitability of each dataset for different research settings.

Our dataset is specifically designed for inverse rendering in a display-camera system. It provides ground truth geometry obtained via scanning, three-dimensional position information for a total of 144 superpixels from an LCD monitor, and stereo views of 16 objects. The stereo camera system is comprised of a polarization camera, which enables the capture of polarization information of the objects.
Figure~\ref{fig:statistics} shows the statistics of our dataset's material.
In dataset described in this paper, objects are placed at 50\,cm from the cameras, we will release additional scene with an object which are placed at multiple distances. 

\section{Additional Details on the Proposed Baseline Model}
In this section, we introduce a simple yet effective baseline model for display-based inverse rendering. The proposed model is designed to handle input images captured under $M$ arbitrary display patterns with intrinsic backlighting, while addressing the challenges posed by limited angular sampling and modeling the effects of near-field lighting.

\paragraph{Initialization}
In the initialization step, we estimate surface normals $\mathbf{n}$ using analytical photometric stereo~\cite{choi2024differentiable}, which operates on $M$ multiplexed images. A depth map is then estimated using RAFT-Stereo~\cite{lipson2021raft}, applied to the average of stereo image pairs captured under multiple patterns. Given this initial geometry, we proceed to optimize the per-pixel reflectance and normal map.

\vspace{-5mm}
\paragraph{Image Formation}
When a scene point is illuminated by a display pattern $\mathcal{L}$, the observed image intensity is modeled as:
\begin{align}
\label{eq:image_formation}
I = \mathrm{clip}\left( \sum_{i=1}^{N} (\mathbf{n}\cdot\mathbf{i}) f(\mathbf{i},\mathbf{o}) 
\frac{L_i}{d_i^2} + \epsilon \right),
\end{align}
where $L_i$ denotes the RGB intensity of the $i$-th superpixel composing the display pattern $\mathcal{L}=\{L_1, \ldots, L_N\}$, $f$ is the BRDF, $\mathbf{n}$ is the surface normal, $\mathbf{i}$ is the incident direction from the $i$-th superpixel, $\mathbf{o}$ is the outgoing view direction, and $d_i$ is the distance between the $i$-th superpixel and the scene point. The function $\mathrm{clip}(\cdot)$ accounts for camera dynamic range clipping, and $\epsilon$ models Gaussian noise.

\paragraph{Simulation for Arbitrary Display Patterns}
To simulate the image under an arbitrary display pattern $\mathcal{P}=\{P_1,\cdots,P_N\}$ based on the equation~\ref{eq:image_formation},
we model the $i$-th display superpixel intensity given the corresponding RGB pattern value we set to display $P_i$ as
\begin{equation}
\label{eq:pattern}
    L_i=s(P_i+B_i)^\gamma,
\end{equation}
where $s$ is a global scalar, $\gamma$ is the non-linear mapping exponent, and $B_i$ is the corresponding spatially-varying backlight intensity.
Then, a scene illuminated by an arbitrary display patten can be described, using Equation~\eqref{eq:image_formation} and Equation~\eqref{eq:pattern}, as:
\begin{align}
\label{eq:relighting}
I\left(\mathcal{P}\right) = \mathrm{clip}\left(\sum_{i=1}^{N} I_i s(P_i+B_i)^\gamma   + \epsilon\right),
\end{align}
where $P_i$ is the display superpixel RGB value, $I_i$ is the captured image under the $i$-th OLAT illumination.
The standard deviation of the Gaussian noise $\epsilon$ can be adjusted to reflect different noise levels.
We represent the captured or rendered image under the $i$-th OLAT illumination, $I_i$ as:
\begin{align}
\label{eq:relighting}
I_i = f(\mathbf{i},\mathbf{o}) (\mathbf{n}\cdot\mathbf{i}) 
\frac{1}{d_i^2}
\end{align}
For modeling near-field effects, spatially-varying lighting direction $\mathbf{i}$ and spatially-varying intensity falloff scalar $\frac{1}{d_i^2}$ are computed by using calibrated superpixel positions and a depth map estimated in initialization step.

\paragraph{Reflectance Model}
Limited angular sampling in display-camera systems presents challenges for accurate reflectance estimation. Following previous approaches~\cite{nam2018practical, chung2024differentiable, li2022neural}, we adopt a basis BRDF representation to regularize the underdetermined nature of the problem. Specifically, we model the SVBRDF as a weighted sum of $J$ basis BRDFs.

Let $w_j$ denote the coefficient for the $j$-th basis BRDF. The overall SVBRDF $f$ is expressed as:
\begin{equation}\label{eq:basis_brdf}
f(\mathbf{i}, \mathbf{o}) = \sum_{j=1}^{J} w_j f_j(\mathbf{i}, \mathbf{o}).
\end{equation}

Each basis BRDF $f_j$ is parameterized by a diffuse albedo $\rho^{d}_{j} \in \mathbb{R}^3$, roughness $\sigma_{j} \in \mathbb{R}^3$, and specular albedo $\rho^{s}_{j} \in \mathbb{R}^3$, using the Cook-Torrance reflectance model~\cite{cook1982reflectance}:
\begin{align}
\label{eq:brdf}
f_j(\mathbf{i}, \mathbf{o}) = \rho^{d}_{j} + \rho^{s}_{j}
\frac{D(\mathbf{h}; \sigma_{j}) F(\mathbf{o}, \mathbf{h}) G(\mathbf{i}, \mathbf{o}, \mathbf{n}; \sigma_{j})}
{4 (\mathbf{n} \cdot \mathbf{i})(\mathbf{n} \cdot \mathbf{o})},
\end{align}
where $\mathbf{h}$ is the half-vector between $\mathbf{i}$ and $\mathbf{o}$, $D$ is the normal distribution function, $F$ is the Fresnel term, and $G$ is the geometric attenuation factor.

\paragraph{Optimization}
At the initialization stage, we performed photometric stereo to estimate surface normals and obtain a pseudo diffuse image. In the HSV color space, we apply K-means clustering on the hue and saturation channels of this pseudo diffuse image to obtain $J$ clusters. Each cluster is converted into a one-hot encoded representation, which serves as an initial estimate of $J$ weight maps. The centroid color of each cluster is assigned as the diffuse albedo of a basis BRDF, initializing the corresponding coefficient. Both roughness and specular albedo are uniformly initialized to 0.5.

We then iteratively optimize the per-pixel surface normals, basis coefficients, and basis BRDFs by minimizing the RMSE loss between the $M$ captured images and the corresponding rendered images. To mitigate noisy per-pixel updates, we incorporate a total-variation (TV) regularization term during optimization.

\section{Additional Experiments}
\begin{figure}[h]
  \centering
    \includegraphics[width=0.55\linewidth]{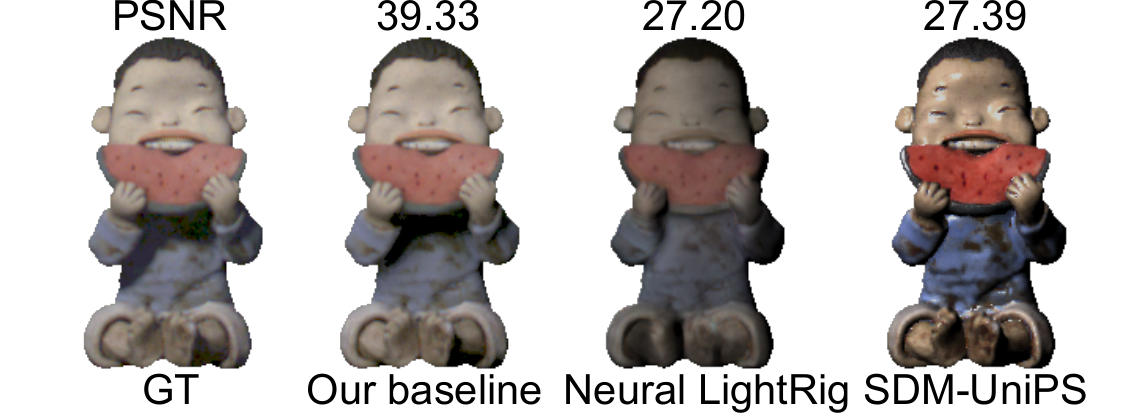}
    \caption{Evaluation of learned methods. }
    \label{fig:learning_based}
\end{figure}

\paragraph{Comparison with Feed-forward Inverse Rendering Models}
Figure~\ref{fig:learning_based} shows additional results for transformer-based SDM-UniPS~\cite{ikehata2023scalable} and diffusion-based Neural LightRig~\cite{he2024neural}. These learned methods show lower-quality results compared to our baseline.
\vspace{-3mm}

\begin{figure}[h]
  \centering
    \includegraphics[width=0.2\linewidth]{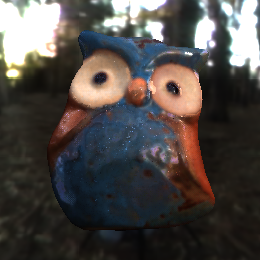}
    \caption{Relighting results with an environment map. }
    \label{fig:env_map}
\end{figure}

\paragraph{Environmental Relighting}
The optimized scene representation enables relighting under a novel environment map as shown in Fig.~\ref{fig:env_map}.

\paragraph{Robustness of Camera Positional Errors}
Our method is robust to camera positional error, obtaining relighting PSNR 38.74 and normal MAE 28.26 for the 5\,cm-displaced camera position. 

\paragraph{Robustness without Stereo Imaging}
Even without stereo imaging, our baseline, using uniform depth, outperforms previous methods, with relighting PSNR 38.8 and normal MAE 28.29, as shown in Table 3. We clarify this explicitly for a fair comparison. 

\begin{figure}[h]
  \centering
    \includegraphics[width=0.65\linewidth]{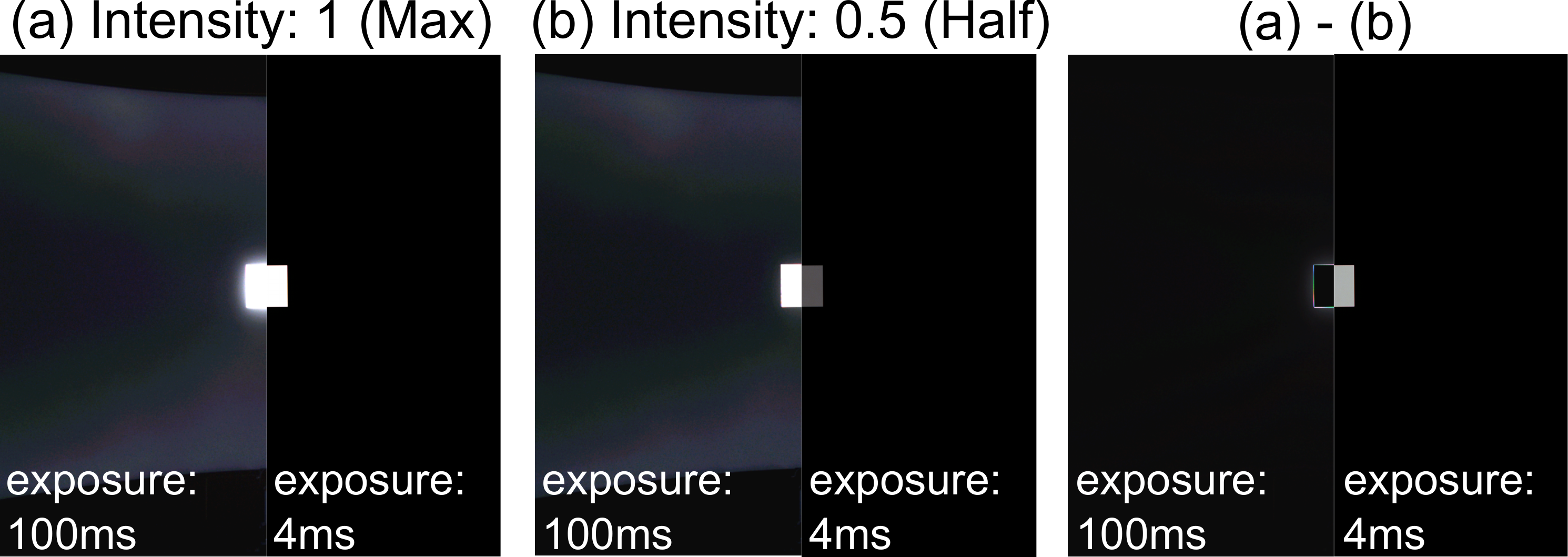}
    \caption{Backlight is invariant to super-pixel intensity: (a) 1 and (b) 0.5. The difference image shows consistent backlight. There is lens flare around the saturated areas.}
    \label{fig:backlight_intensity}
\end{figure}

\paragraph{Analysis of Display Backlight}
Figure~\ref{fig:backlight_intensity} (a) and (b) show images of the display when the superpixel intensity is set to 1 (maximum) and 0.5 (half), respectively. 
The difference image shows that the backlight remains constant with different intensities in signal inputs.

\begin{figure}[h]
\vspace{-0mm}
  \centering
    \includegraphics[width=0.7\linewidth]{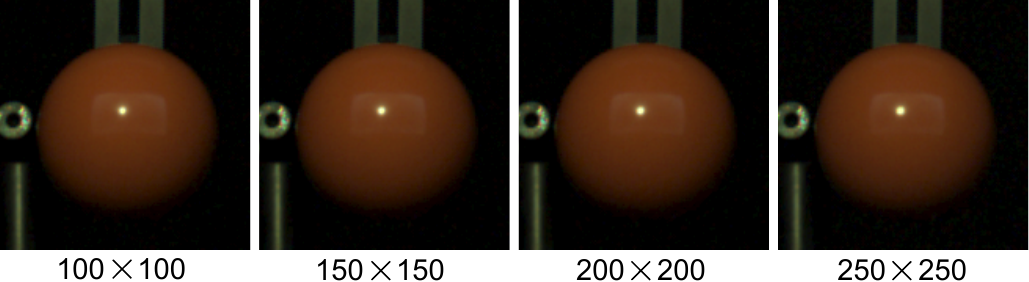}
    \caption{
    Images of a glossy sphere with superpixels of different sizes. Numbers below indicate pixels per superpixel.}
    \label{fig:pointlight}
\end{figure}

\paragraph{Analysis of Point-light Assumption as a Superpixel}
Assuming a superpixel as a point light source introduces a trade-off between blurred reflectance and noise caused by limited exposure. 
Although the design choice in the baseline deviates from the ideal point light assumption, Figure~\ref{fig:pointlight} demonstrates that the size of the superpixel has a minimal impact on the specular lobe.

\section{Photometric Stereo Results}
The following are the surface normal reconstruction results for all the photometric stereo methods we evaluated.

\begin{figure*}[t]
	\centering
		\includegraphics[height=0.9\textheight]{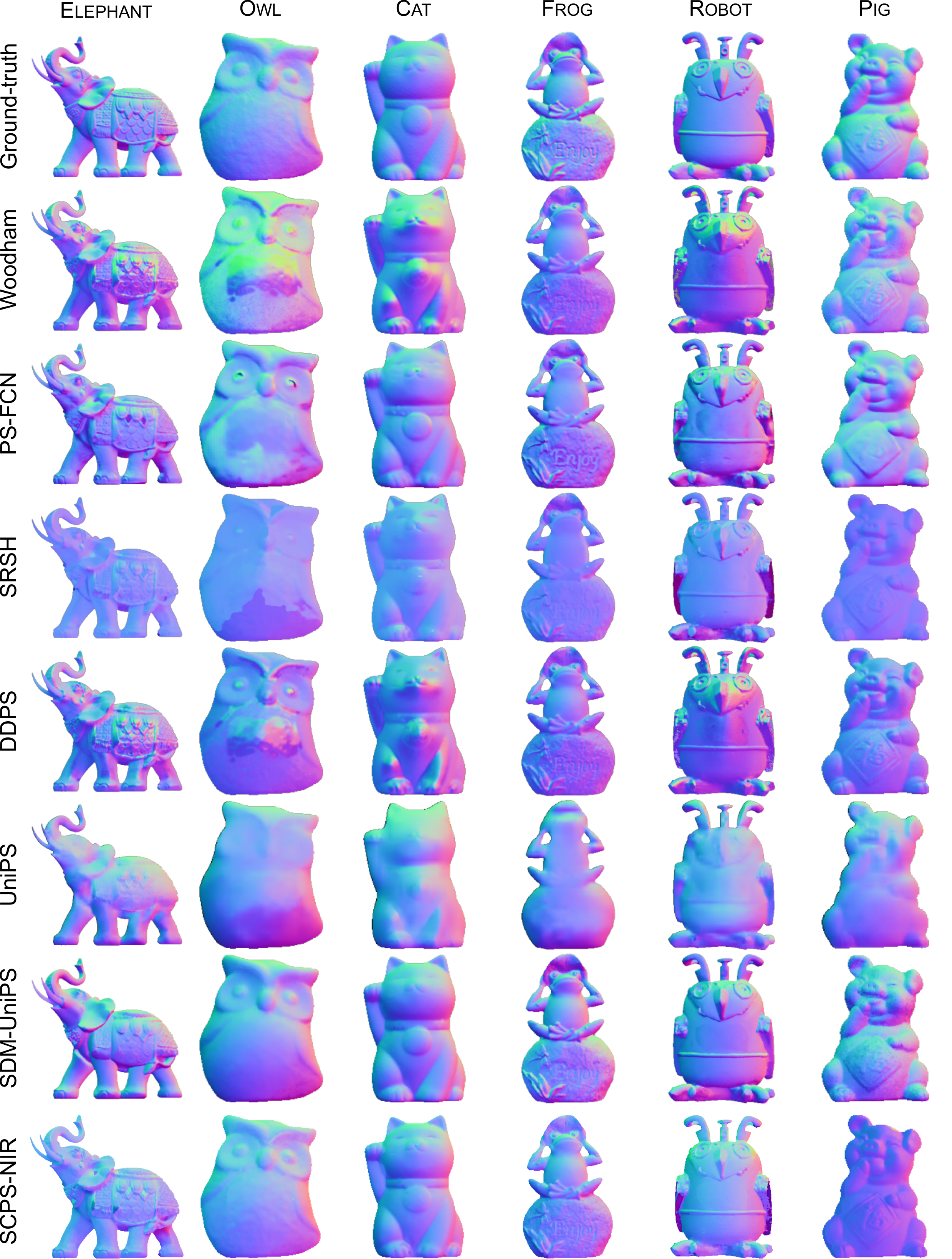}
		\caption{\textbf{Reconstructed normal of evaluated methods.} 
        We visualize reconstructed normals of \textsc{Elephant}, \textsc{Owl}, \textsc{Cat}, \textsc{Frog}, \textsc{Robot}, \textsc{Pig}}
  		\label{fig:all_ps1}
\end{figure*}

\begin{figure*}[t]
	\centering
		\includegraphics[height=0.9\textheight]{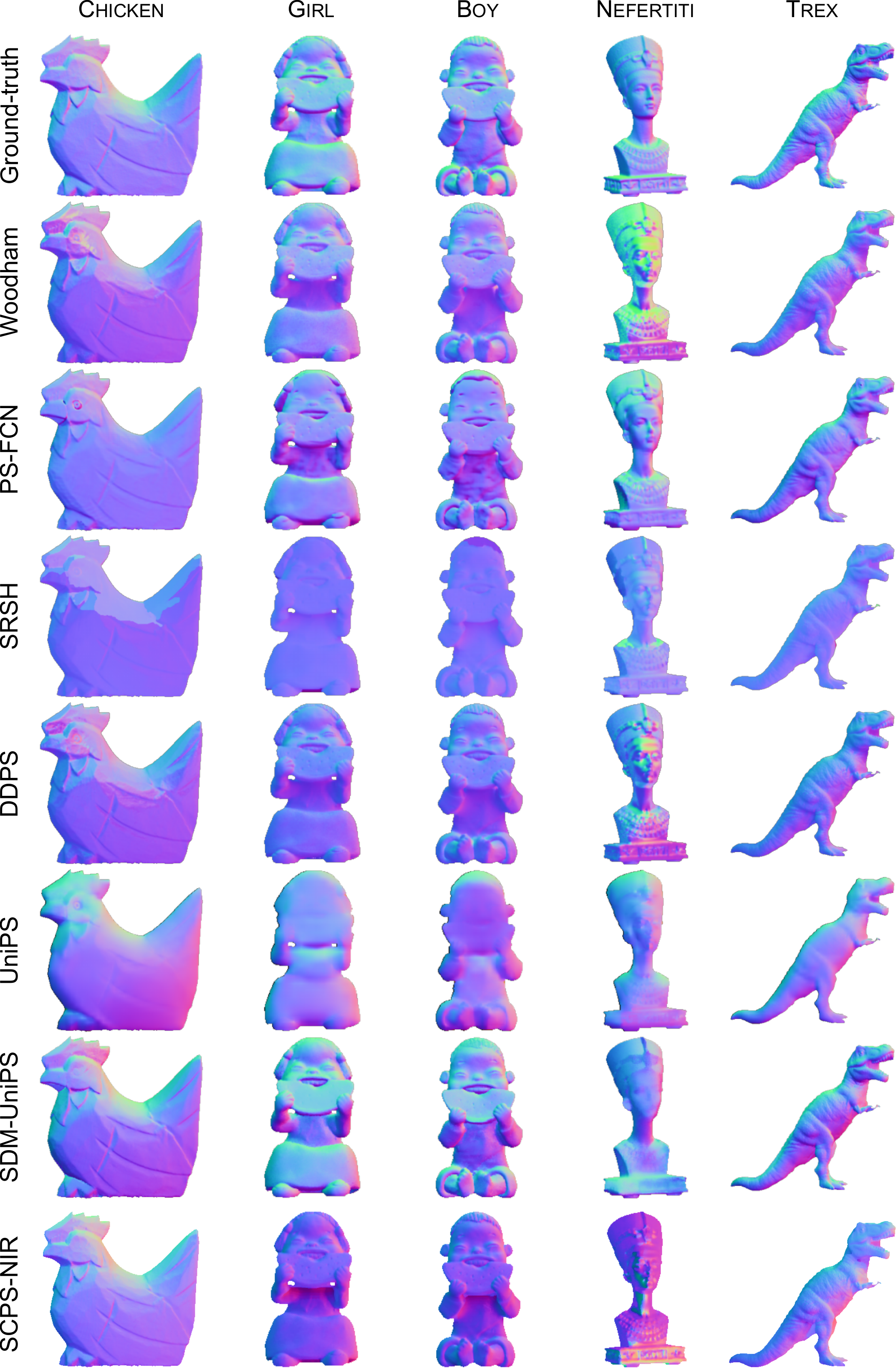}
		\caption{\textbf{Reconstructed normal of evaluated methods.}
        We visualize reconstructed normals of \textsc{Chicken}, \textsc{Girl}  \textsc{Boy}, \textsc{Nefertiti}, \textsc{Trex}}
  		\label{fig:all_ps2}
\end{figure*}

\begin{figure*}[t]
	\centering
		\includegraphics[height=0.9\textheight]{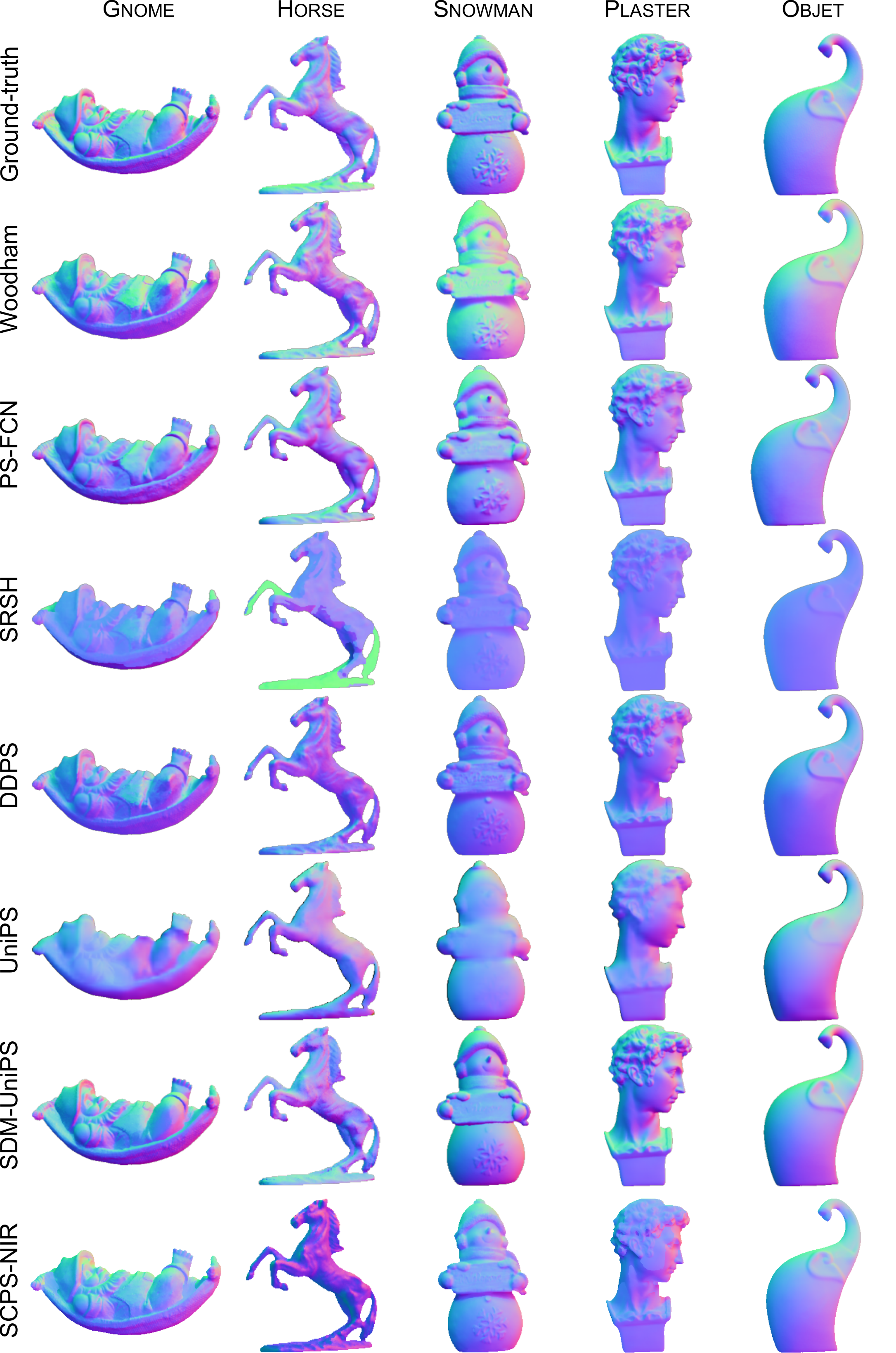}
		\caption{\textbf{Reconstructed normal of evaluated methods.}
        We visualize reconstructed normals of \textsc{Gnome}  \textsc{Horse}, \textsc{Snowman}, \textsc{Plaster}, \textsc{Objet} }
  		\label{fig:all_ps3}
\end{figure*}

\clearpage

{
    \small
    \bibliographystyle{ieeenat_fullname}
    \bibliography{references}
}
\end{CJK}